%% file: ms.tex
\documentclass[12pt,flushrt, preprint]{aastex}

\usepackage{amsmath}

\newcommand{\pa}[1]{\left(#1\right)}
\newcommand{\pac}[1]{\left[#1\right]}

\newcommand{\abs}[1]{\left|#1\right|}
\newcommand{\norme}[1]{\Arrowvert #1 \Arrowvert}

\newcommand{\Rep}[1]{\Re\pa{#1}}
\newcommand{\Imp}[1]{\Im\pa{#1}}

\newcommand{\vect}[1]{\overrightarrow{#1}}

\newcommand{\plus}{\mbox{\raisebox{.2mm}{\tiny{\ensuremath{+}}}}}
\newcommand{\moins}{\mbox{\raisebox{.2mm}{\tiny{\ensuremath{-}}}}}
\newcommand{\pmoins}{\mbox{\raisebox{.2mm}{\tiny{\ensuremath{\pm}}}}}

\newcommand{\ddf}[2]{\frac{\dd#1}{\dd#2}}

\newcommand{\ddxx}[2]{\frac{\dd^2{}#1}{\dd{}{#2}^2}}
\newcommand{\ddtt}[1]{\ddxx{#1}{t}}

\newcommand{\DDp}[2]{\frac{\partial^2#1}{\partial{#2}^2}}
\newcommand{\Dpc}[3]{\frac{\partial^2#1}{\partial#2\,\partial#3}}

\newcommand{\cf}{\emph{cf} }
\newcommand{\ttt}[1]{\texttt{#1}}
\newcommand{\dS}{dS}

\newcommand{\MW}{Milky Way}
\newcommand{\marge}{\noindent}

\newcommand{\f}[2]{\frac{#1}{#2}}

\newcommand{\tf}[2]{{#1}/{#2}}
\newcommand{\patf}[2]{\pa{{#1}/{#2}}}

\newcommand{\e}[1]{_{\mathrm{#1}}}
\newcommand{\dd}{\mathrm{d}}
\newcommand{\ph}{\varphi}
\newcommand{\eps}{\varepsilon}

\newcommand{\U}[1]{\ensuremath{\mathrm{~#1}}}

\begin{document}


\title{Parametric Dwarf Spheroidal Tidal Interaction}

\author{Jean-Julien Fleck\altaffilmark{1}}
\affil{Institute for Astronomy, University of Hawaii, 2680 Woodlawn Dr., Honolulu, HI, 96822, USA}

\altaffiltext{1}{Permanent address:
D\'epartement de Physique de l'\'Ecole
Normale Sup\'erieure, 24 rue Lhomond, F-75231
Paris Cedex 05, France}
\email{jfleck@ifa.hawaii.edu}

\and

\author{J. R. Kuhn}
\affil{Institute for Astronomy, University of Hawaii, 2680 Woodlawn Dr., Honolulu, HI, 96822, USA}
\email{kuhn@ifa.hawaii.edu}


\input{abstract.tex}



\input{introduction.tex}


\input{parametric.tex}


\input{numeric.tex}


\input{observed_comparisons.tex}


\input{conclusions.tex}


\input{bibliography.tex}


\input{PS/tide.tex}


\input{PS/rkzero.tex}

\input{PS/rkinf.tex}

\input{PS/rk_1.tex}

\input{PS/ellrk.tex}


\input{PS/eps_fft_bon.tex}


\input{PS/relax_dtime_bon.tex}


\input{PS/omega_snapshot_bon.tex}


\input{PS/omegaomega_bon.tex}


\input{PS/100by100_ejection_bon.tex}


\input{PS/veldisp.tex}

\input{PS/dsvelpic.tex}


\input{PS/Film.tex}


\input{PS/order_Var_bon.tex}


\input{PS/dsprops.tex}

\end{document}

%% file: abstract.tex
\begin{abstract}

The time dependent tidal interaction of the Local Group Dwarf
Spheroidal (dS) Galaxies with the Milky Way (MW) can fundamentally
affect their dynamical properties. The model developed here extends
earlier numerical descriptions of dS-MW tidal interactions.  We
explore the dynamical evolution of dS systems in circular or elliptical
MW orbits in the framework of a parametric oscillator.  An analytic
model is developed and compared with more general numerical solutions
and N-body simulation experiments.

\end{abstract}

%% file: introduction.tex
\section{Introduction}

The Local Group dwarf spheroidal (dS) galaxies may be sensitive probes
of the chemical and dynamical formation history of the Milky Way (\cf 
Harbeck et al. 2001). They have also been used to constrain dark
matter models and early structure formation scenarios in the Universe (\cf
{\L}okas 2002).

Controversy over whether the dS are equilibrium dynamical systems (\cf
Mateo 1998) or tidally driven non-equilibrium objects (Kuhn and
Miller 1989, Kroupa 1997) persists. The primary observational evidence for tidally
dominated dynamics comes not only from the dS high velocity dispersions and highly elliptical shapes but also from measurements of a spatially variable
velocity dispersion (Kleyna et al. 2002)  and significant stellar
populations beyond the dS nominal tidal radii (Kuhn et al. 1996,
Mart{\'{\i}}nez-Delgado et al. 2001). While there is undisputed
evidence that the dS in Sagitarius is tidally disrupted (Ibata et al. 1994)
some practitioneers have devised more complex multiparametric models
which could formally account for the kinematic data in some of the
high M/L dwarfs using dark matter (Kleyna et al. 2002, Wilkinson et al. 2002, Lokas 2002).

A theme of the tidal excitation argument presented here is that the
gravitational interaction scale for transfering energy to the dS
stellar system from the dS-MW orbit can be much larger than the dS
classical tidal radius.  In those systems where the stellar crossing
time can be comparable to the ``period'' of the external tide, the
simple static arguments for equilibrium break down. Early calculations
(Kuhn and Miller 1989 --  henceforth KM) demonstrated this possibility
using particle-mesh gravitational N-body simulations in idealized
circular dS orbits.

This paper develops an analytic formulation of the dS-MW tidal
interaction problem. Once again, by treating the response of the dS in
terms of galaxy oscillations, but now incorporating the
tidal interaction through the Mathieu equation, it is possible to
understand the evolution of these tidally interacting systems in
non-circular orbits far from the simple resonance condition.
As a parametric oscillator we find that a dS stellar system can
be ``inflated'' even when its characteristic
pulsational spectrum is not tuned to its MW orbital dynamics.










%% file: parametric.tex
\section{Parametric tidal interactions}

An oscillator driven by a periodic forcing function can sometimes be described
by the Mathieu equation

\begin{equation}
\ddot{x} + {\omega_0}^2 \pa{1 - \eps\, \cos{2\omega t}} x		
\label{Mat:simp}
		 = 	0
\end{equation}

\marge Here $x(t)$ describes the oscillator amplitude function, $\omega_0$
is the oscillator resonant frequency, $\omega$ is the driving frequency
and $\eps$ is a small constant. Solutions to equation (\ref{Mat:simp}) are
most easily developed by a perturbative multi-timescale analysis (Bender
and Orzag 1978). One finds that, in general, growing (unstable)  
oscillatory solutions exist for $\omega = \omega_0/n$, $n \in \mathbb{N}$.

An important property of equation (\ref{Mat:simp}) is that even for a
weakly damped (high `Q') oscillator a significant, non-zero, frequency range
for the driving force leads to growing unstable
solutions.  For small intrinsic damping this result is independent of
the frequency width of the natural resonance. For example, near the
first resonance $\omega$ between
$\omega_0\pa{1-\eps /4}$ and $\omega_0\pa{1+\eps /4}$ 
leads to instability.  The second resonance near
$\omega=\omega_0/2$ is unstable for a narrower frequency range
\begin{equation}
\f{1}{24}\eps^2\omega ~>~ \omega-\omega_0/2 ~>~ \f{-5}{24}\eps^2\omega
\end{equation}

We show here that the internal dynamics of stellar systems
in the MW's gravitational environment can be described as a two-dimensional
coupled parametric oscillator. These equations can be solved analytically
using multi-scale perturbation methods to explain the dynamical
and morphological properties of the Local Group dS.

\subsection{Tide geometry}

Each of the dS orbit the MW at a distance which is long compared to the
galactic disk scale length so that it is reasonable to treat
the MW as a purely spherical central force problem. We define the MW
central force

\begin{equation}
\vect{F}\e{\!MW} = \vect{F}(\vect{r}) = F(r)\, \widehat{r}
\end{equation}

\marge	 Letting $\vect{r}$ describe the displacement between the
MW and dS centers and $\vect{\delta r}$ be the displacement to a dS star from
the dS center we write the MW tidal force on the star at lowest order as 

\begin{eqnarray}
\vect{\Delta F}\e{\!\!MW}	
		& = &	\vect{F}\pa{\vect{r} + \vect{\delta r}} - 
			\vect{F}\pa{\vect{r}} 	\nonumber	\\
		& = &	\pac{F\pa{\norme{\vect{r} + \vect{\delta r}}} - 
			     F\pa{\norme{\vect{r}}}}\, 
			\widehat{r}			\nonumber	\\
		&   &	+ \sin{\pa{\Delta\ph}}\, 
			F\pa{\norme{\vect{r}}}\, 
			\widehat{\ph} \label{tide:exact}
\end{eqnarray}

\marge Here $\Delta\ph$ is the central angle between a dS star, the MW,
and the
dS center and $\widehat{r}$ and $\widehat{\ph}$ are unit vectors in the
radial and central angle directions (Figure~\ref{fig:tide}).
Since the spatial extent of the dS is small compared to
its distance from the galactic center we expand equation
(\ref{tide:exact}) using

\begin{mathletters}
\begin{eqnarray}
F\pa{\norme{\vect{r} + \vect{\delta r}}} -
F\pa{\norme{\vect{r}}}
			& = &	\ddf{F}{r}(r)\, 
				\vect{\delta r}\cdot \widehat{r}	\\
\sin{\pa{\Delta\ph}}	& = &	\f{1}{r}\, 
				\vect{\delta r}\cdot \widehat{\ph}	
\end{eqnarray}
\end{mathletters}

\marge We define a cartesian ($xy$) coordinate system centered on the dS
which contains the MW and the star so that we can express $\vect{\delta
r}$, $\widehat{r}$ and $\widehat{\ph}$ in term of $\widehat{x}$ and
$\widehat{y}$.

\begin{mathletters}
\begin{eqnarray}
\vect{\delta r}		& = &	x\, \widehat{x} + y\, \widehat{y}	\\
\widehat{r}	& = &	\cos\ph \, \widehat{x} +
				\sin\ph \, \widehat{y} \label{def:hat:r}\\
\widehat{\ph}	& = &	-\sin\ph \, \widehat{x} + 
				\cos\ph \, \widehat{y} \label{def:hat:phi}	
\end{eqnarray}
\end{mathletters}

\marge  The MW orbital time dependence is contained in $r$ and $\ph$.
To this order of approximation these equations also describe the
three-dimensional stellar problem where $(x,y)$ are the projected
coordinates in the dS-MW orbital plane. The tidal force perpendicular
to the orbital plane yields harmonic motion which is not coupled to
the  MW orbital motion and is ignorable.  Thus the tidal
force becomes,

\begin{mathletters}
\begin{eqnarray}
\Delta F_x	& = &	\ddf{F}{r}(r)
			\pa{x\, \cos^2\ph + \f{y}{2}\, \sin{2\ph}}
			+ \f{F(r)}{r}
			\pa{x\, \sin^2\ph - \f{y}{2}\, \sin{2\ph}}
									\\
\Delta F_y	& = &	\ddf{F}{r}(r)
			\pa{y\, \sin^2\ph + \f{x}{2}\, \sin{2\ph}}
			+ \f{F(r)}{r}
			\pa{y\, \cos^2\ph - \f{x}{2}\, \sin{2\ph}}
\end{eqnarray}
\end{mathletters}

The form of the MW potential determines $\ddf{F}{r}$ and
$F$. For a Keplerian potential, it is \mbox{$\ddf{F}{r} = -2\,
\f{F(r)}{r}$} and for the logarithmic case it is \mbox{$\ddf{F}{r} =
-\f{F(r)}{r}$}. Hence we can write a simplified form of the tide for these
two potential forms:

\begin{mathletters}
\begin{eqnarray}
\vect{\Delta F}\e{Kep}	
		& = & 	\f{3}{2}\ddf{F}{r} \left[ \vphantom{\f{1}{2}}
			 \pa{x\,\cos{2\ph} - \f{x}{3}+
			     y\,\sin{2\ph}} \widehat{x} 
			\right.						\\
		&   &	\left. \hskip 2cm \vphantom{\f{1}{2}}
			+\pa{-y\,\cos{2\ph} - \f{y}{3} +
			      x\,\sin{2\ph}} \widehat{y} 
				\right]				\nonumber	\\
\vect{\Delta F}\e{log}
		& = & 	\ddf{F}{r} \left[ \vphantom{\f{1}{2}}
			 \pa{ x\,\cos{2\ph} + y\,\sin{2\ph}} \widehat{x} 
			\right.						\\
		&   &	\left. \hskip 2cm \vphantom{\f{1}{2}}
			+\pa{-y\,\cos{2\ph} + x\,\sin{2\ph}} \widehat{y}
			\right]				\nonumber
\end{eqnarray}
\end{mathletters}

\marge	The numerical factors are of course different for different
potential assumptions but the form of the resonant solution remains
unchanged. It seems likely that the MW potential is actually
logarithmic in the domain of most dS so that we assume the form
\begin{equation}
\phi = {v_c}^2\, \ln\pa{\f{r}{r_0}}
\end{equation}

\subsection{Solution of the coupled Mathieu equations at first order}

\subsubsection{Equation of motion for a star near the core}

In the dS reference frame a star experiences forces due to the MW tide and
the dS potential. The dS potential generates a radial acceleration,
$a_{r}= -GM(r)/r^2$, where $M(r)$ is the mass enclosed within a spherical
volume of radius $r$ centered on the dS. Near the center of the dS we
approximate $M(r)\approx \rho_0 4\pi r^3/3$ so that we can write
$a_{r}\approx -{\omega_0}^2 r$ with ${\omega_0}^2=4\pi \,G\rho_0/3$.  In
this case the total dS-frame acceleration of a star can be written as

\begin{equation}
\ddtt{\vect{r}} = 	-{\omega_0}^2\, \vect{r} + 
				\f{\vect{\Delta F}}{m}
\end{equation}

\marge	With additional definitions,

\begin{mathletters}
\begin{eqnarray}
k		& = &	\f{1}{m}\ddf{F}{r}(r)		\label{def:k}	\\
\eps		& = &	\f{k}{{\omega_0}^2}~
			=~\f{{\omega_c}^2}{{\omega_0}^2}
\end{eqnarray}
\end{mathletters}

\marge	where the circular frequency of the \dS{} around the \MW{} 
is $\omega_c=v_c/r$. Combining
these terms we finally obtain a 
recognizable set of 
coupled equations that describe the
behavior of a star in the center of a dS as projected in the $(xy)$ plane.

\begin{mathletters}\label{eq:xy}
\begin{eqnarray}
\ddot{x} + {\omega_0}^2 \pa{1 - \eps\, \cos{2\ph}} x		\label{Mat:gen:x}
		& = &	\eps\, {\omega_0}^2\,  \sin{2\ph}\: y		\\
\ddot{y} + {\omega_0}^2 \pa{1 + \eps\, \cos{2\ph}} y		\label{Mat:gen:y}
		& = &	\eps\, {\omega_0}^2\,  \sin{2\ph}\: x
\end{eqnarray}
\end{mathletters}

Notice that the coupling between $x$ and $y$ motion results
from the anisotropic form of the galactic tide where
the expansive and compressive tidal contributions are both important in
determining the dS dynamical response.

\subsubsection{Galaxy oscillations}

Individual stars also experience forces due to the dS gravitational {\it
perturbations} but equations (\ref{eq:xy}) neglect temporal 
variations
(due to $\rho (t)$) in the dS potential.  Thus there is an additional acceleration
term which should be included here at late times in the dS-tide evolution
when the temporal variability of the dS density is important.

The resulting oscillations in the mean stellar orbits are not included above
but can be described in a manner which is analogous to
polytropic stellar normal mode oscillations. This description of galaxy
dynamics in terms of linear oscillations was the basis of KMs prediction
that Local Group dS could be tidally inflated even when the stars were
contained within the static tidal radius. Miller
and Smith (1994) and VanderVoort (1999) have now demonstrated that
weakly damped or even growing finite amplitude normal mode galaxy
oscillations are a physically interesting feature of realistic galaxy
dynamics simulations. 

The lowest order oscillation mode should have no radial nodes -- as is the
case for radial oscillations in stellar polytropes (\cf Cox 1980). VanderVoort
(1999) showed that the total Lagrangian radial displacement summed over
all stars in an N-body system exhibits a very similar oscillatory
behavior. This is not entirely intuitive given that the collisionless
orbital motion of stars in a galaxy occurs over the length scale of the
galaxy, while a mass element in a stellar polytrope oscillates only over the
small scale of the oscillation displacement amplitude.
Nevertheless VanderVoort proves that the fundamental N-body radial
eigenmode for the mean Lagrangian radial particle displacement,
$\vect{\xi}$, satisfies

\begin{equation} 
\ddtt{\vect{\xi }} = -{\omega_0}^2\vect{\xi} \label{oscfreq} 
\end{equation} 

\marge with 
 ${\omega_0}^2=-W/I=G\int \!M(r)\rho(r)\, r\,\dd r /\!\int\!
\rho(r) \,r^4\,\dd r$ and $\vect{\xi} \propto \vect{r}$. Here $W$ and $I$
are the potential energy and moment of inertia, while $\rho (r)$ and
$M(r)$ are the galaxy density and enclosed mass distribution and $G$ is
the gravitational constant. Realistic density distributions lead to
frequencies that are of order $\pi\sqrt{G\rho_0}$ where $\rho_0$ is the
central density.

\subsubsection{Circular orbit}
\label{sect:circular}

If we launch the dS on a circular orbit around the MW, then $\ph=\omega t$
and equations (\ref{eq:xy}) describe a set of two-dimensional Mathieu
equations.  We now show that these equations exhibit resonant instability
around \mbox{$\omega = \omega_0$}. To obtain a valid solution near the
resonance, we use a multi-timescale analysis in terms of $t$ and
\mbox{$\tau = \eps t$} and develop the relevant variables in terms of the
perturbation parameter $\eps$.

\begin{mathletters}
\begin{eqnarray}
\omega_0 	& = & 	\omega + \eps\, \omega_1			\\
x(t,\tau)	& = &	x_0(t,\tau) + \eps\, x_1(t)			\\
y(t,\tau)	& = &	y_0(t,\tau) + \eps\, y_1(t)	
\end{eqnarray}
\end{mathletters}

The equations obtained from the lowest order ($\eps^0$) for $x_0$ and $y_0$ are
simple harmonic oscillators with frequency $\omega_0$  as we would expect in a harmonic system. From the lowest order $x_0$ and $y_0$
solutions and first order equation we can construct a refined solution.
Using complex notation we write

\begin{mathletters}
\begin{eqnarray}
x_0(t,\tau)	& = &	A(\tau)\, \text{e}^{\text{i} \omega t} + \text{c.c.}									\\
y_0(t,\tau)	& = &	B(\tau)\, \text{e}^{\text{i} \omega t} + \text{c.c.}
\end{eqnarray}
\end{mathletters}

\marge	Gathering the terms of order $\eps^1$, we obtain differential equations
for $x_1$ and $y_1$

\begin{mathletters}
\begin{eqnarray}
\DDp{x_1}{t} + {\omega}^2\, x_1
		& = &	-2\omega \omega_1\, x_0
                        +{\omega}^2\,\cos{\pa{2\, \omega t}}\, x_0
                        +{\omega}^2\,\sin{\pa{2\, \omega t}}\, y_0
                        -2\, \Dpc{x_0}{t}{\tau} \label{x1eqn}         \\
\DDp{y_1}{t} + {\omega}^2\, y_1
		& = &	-2\omega \omega_1\, y_0
                        -{\omega}^2\,\cos{\pa{2\, \omega t}}\, y_0
                        +{\omega}^2\,\sin{\pa{2\, \omega t}}\, x_0
                        -2\, \Dpc{y_0}{t}{\tau} \label{y1eqn}
\end{eqnarray}
\end{mathletters}

\marge These are combined with the lowest order solutions for $x_0$ and
$y_0$. We construct a solution for $A(\tau)$ and $B(\tau)$ using the
well-known ansatz (Fredholm alternative) of forcing all harmonic terms
which lead to unstable (secular) solutions to vanish. Obtaining a solution
with unbounded coefficients implies that there is no stable oscillatory
solution for $x$ and $y$. Thus we force the coefficients of
$\text{e}^{\text{i} \omega t}$ and $\text{e}^{-\text{i} \omega t}$ on the
RHS of equations (\ref{x1eqn}) and (\ref{y1eqn}) to vanish to obtain two
differential equations (and their conjugates) for the functions $A(\tau)$
and $B(\tau)$:

\begin{mathletters}
\begin{eqnarray} 
-2\, \omega_1\, A(\tau) + \f{\omega}{2}\, A^*(\tau)
        +\f{\omega}{2\, \text{i}}\, B^*(\tau)
        - 2\, \text{i}\, \ddf{A(\tau)}{\tau}
		& = &	0				\\
-2\, \omega_1\, B(\tau) - \f{\omega}{2}\, B^*(\tau)
        +\f{\omega}{2\, \text{i}}\, A^*(\tau)
        - 2\, \text{i}\, \ddf{B(\tau)}{\tau}
		& = & 0
\end{eqnarray}   
\end{mathletters}

	We solve these four coupled differential equations in terms of the real and
imaginary parts of $A$ and $B$, i.e. $(\Rep{A}, \Imp{A}, \Rep{B}, \Imp{B})$. This yields the matrix equation

\begin{equation}
\ddf{X}{\tau} = M\, X \label{Mat:eigen}
\end{equation}
where,
\begin{equation}
X 	 =  	\begin{pmatrix}
			\Rep{A}  \\
			\Imp{A}  \\
			\Rep{B}  \\
			\Imp{B}
		\end{pmatrix}
\end{equation}  
\begin{equation}
        M = \f{1}{4}
            \begin{pmatrix}
                0       &       -\pa{4\,\omega_1 + \omega}      &
                -\omega &       0       \\
                \pa{4\,\omega_1 - \omega}       &       0       &
                0       &       \omega  \\
                -\omega &       0       &
                0       &       - \pa{4\,\omega_1 - \omega}     \\
                0       &       \omega  &
                \pa{4\,\omega_1 + \omega}       &       0       \\
            \end{pmatrix}
\end{equation}

\noindent      This equation is solved most readily by finding the 
eigenvectors of $M$. Taking

\begin{equation}
H=\sqrt{\f{\omega+2\,\omega_1}{\omega-2\,\omega_1}}	\label{def:H}
\end{equation}

\noindent the eigenvectors of M of respective eigenvalues
$\f{\sqrt{\omega^2-4\,{\omega_1}^2}}{2}$,
$-\f{\sqrt{\omega^2-4\,{\omega_1}^2}}{2}$,
$\text{i}\, \omega_1$ and $-\text{i}\, \omega_1$ are

\begin{equation}
        \begin{pmatrix}
	-H\\1\\1\\H
        \end{pmatrix},
        \begin{pmatrix}
	H\\1\\1\\-H
        \end{pmatrix},
        \begin{pmatrix}
        1\\-\text{i}\\\text{i}\\1
        \end{pmatrix},
        \quad \text{and} \quad 
        \begin{pmatrix}
        1\\\text{i}\\-\text{i}\\1
        \end{pmatrix}
\end{equation}

\marge The solution to equation (\ref{Mat:eigen}) is obtained in terms of 
four
complex constants of integration $C=(\alpha,\beta,\gamma,\delta)$ and
the eigenvectors as:

\begin{mathletters}
\begin{eqnarray}
        \alpha\,
        \exp\pac{\f{\sqrt{\omega^2 - 4\,{\omega_1}^2}}{2}\,\tau}
                & = &	-H\, \Rep{A} + \Imp{A} +
                        \Rep{B} +  H\, \Imp{B}    \\
        \beta\,
        \exp\pac{-\f{\sqrt{\omega^2 - 4\,{\omega_1}^2}}{2}\,\tau}
                & = &	\hphantom{-}
                        H\, \Rep{A} + \Imp{A} +
                        \Rep{B} - H\, \Imp{B}     \\
        \gamma\,\text{e}^{\text{i}\omega_1\tau}
                & = &	\hphantom{-H\,}
                        \Rep{A} - \text{i}\,\Imp{A} +
                        \text{i}\,\Rep{B} + \Imp{B}       \\
        \delta\,\text{e}^{-\text{i}\omega_1\tau}
                & = &	\hphantom{-H\,}
                        \Rep{A} + \text{i}\,\Imp{A} -
                        \text{i}\,\Rep{B} + \Imp{B}
\end{eqnarray}
\end{mathletters}

	At late times the growing exponential term dominates, and these equations
simplify to yield expressions for $A$ and $B$,

\begin{mathletters}
\begin{eqnarray}
A(\tau)	& = &	\f{\alpha}{4\, H}
		\exp\pac{\f{\sqrt{\omega^2 - 4\,{\omega_1}^2}}{2}\,\eps t}
		\pa{-1 + \text{i}\, H}					\\
B(\tau)	& = &	\f{\alpha}{4\, H}
		\exp\pac{\f{\sqrt{\omega^2 - 4\,{\omega_1}^2}}{2}\,\eps t}
		\pa{H + \text{i}}
\end{eqnarray}
\end{mathletters}

\marge Evidently $A$ and $B$ define the mean position of a star in
the dS. 
\begin{mathletters}\label{eq:xygen}
\begin{eqnarray}
x(t)	& = &	\alpha'\,
		\exp\pac{\f{\sqrt{\omega^2 - 4\,{\omega_1}^2}}{2}\,\eps t}
		\cos\pa{\omega t + \theta}	\label{xvstime}	\\
y(t)	& = &	\alpha'\,
		\exp\pac{\f{\sqrt{\omega^2 - 4\,{\omega_1}^2}}{2}\,\eps t}
		\sin\pa{\omega t + \theta} 	\label{yvstime}
\end{eqnarray}
\end{mathletters}

\marge Equations (\ref{xvstime}) and (\ref{yvstime}) show that 
instability 
occurs if
$\omega > \abs{2\omega_1}$. Thus if the circular frequency
of the orbit satisfies 

\begin{equation}
\abs{\omega_0-\omega_c} < \omega_c\,\eps/2 \label{eq:width}
\end{equation}
the dS will be tidally excited. 
The constants $C$ determine the initial position and
velocity of a given star in the dS. Notice that the growing eigenmode solution
involves stars oriented along  a line that makes an angle $\theta$
with respect to the $x$ axis where $\tan\theta = -H.$
Thus, we find in general that stars expand
away from
the center of the dS depending on the difference
between the dS orbital frequency and $\omega_0$, and on the perpendicular 
distance
a star makes to a line in the $(xy)$ plane of the dS at angle $\theta$ 
with 
respect to the line-of-sight toward the MW. 

Extending this calculation to second order in $\eps$ reveals
that, unlike the simple one-dimensional case,
the next resonance near $\omega_0/2$ is stable with no
secular amplitude variation.

\subsubsection{Elliptical orbit}

Equations (\ref{xvstime}) and (\ref{yvstime}) assume a circular orbit.  
Their applicability is also restricted, depending upon the central
concentration and oscillation frequency $\omega_0$ of the dS. For example,
the perturbation parameter, $\eps = {\omega_c}^2/{\omega_0}^2$, can
approach unity for a circular orbit in a logarithmic potential. Under
these conditions the static tide force alone is strong enough to disrupt the dS. Since the
dS are generally not in circular orbits it is important to extend the model
to elliptical orbits. In a logarithmic MW potential we can recover the form of the Mathieu equation with the epicycle approximation. We define an
``eccentricity'' $e$ from $r(t)$ where 
\mbox{$\kappa = \sqrt{2}\, \omega$} is the epicycle frequency for a
logarithmic potential. It follows that (Binney and Tremaine 1987),

\begin{mathletters}
\begin{eqnarray}
r(t) 	& = &	r_c \pa{1 - e\, \cos\pa{\kappa t}}	\\
\ph(t)	& = &	\omega_c t +  e\,\sqrt{2}\, \sin\pa{\kappa t}
\end{eqnarray}
\end{mathletters}

\marge Here \mbox{$k = \patf{v_c}{r(t)}^2\approx{\omega_c}^2 \pa{1 + 2e\,
\cos\pa{\kappa t}}$} where \mbox{$v_c = r_c\,\omega_c$} is the circular
velocity of the potential. We note that with this convention for $r(t)$,
the \dS{} is at perigalacticon at $t=0$.  Replacing \mbox{$k(t) =
\eps(t)\, {\omega_0}^2$} and $\ph(t)$ in the equations (\ref{eq:xy})
gives, at first order in $e$,

\begin{mathletters}\label{eq:xy:ellips}
\begin{eqnarray}
\ddot{x} + {\omega_0}^2 \pa{1 - \eps_c \cos\pa{2 \omega_c t} -
\sum_{\pmoins}
b_{\pmoins}\, e\, \eps_c \cos\pa{2a_{\pmoins}\,\omega_c t}} x
                \ = \   \hskip 4cm \nonumber \\
{\omega_0}^2\pa{\eps_c \sin\pa{2 \omega_c t} +
        \sum_{\pmoins}
                                \label{xellips}
b_{\pmoins}\, e\, \eps_c \sin\pa{2a_{\pmoins}\,\omega_c t}} y
\\
\ddot{y} + {\omega_0}^2 \pa{1 + \eps_c \cos\pa{2 \omega_c t} +
\sum_{\pmoins}
        b_{\pmoins}\, e\, \eps_c \cos\pa{2a_{\pmoins}\,\omega_c t}} y
                \ = \ \hskip 4cm \nonumber      \\
{\omega_0}^2\pa{\eps_c \sin\pa{2 \omega_c t} +
                                \label{yellips}
\sum_{\pmoins}b_{\pmoins}\, e\, \eps_c\sin\pa{2a_{\pmoins}\,\omega_c t}}x
\end{eqnarray}
\end{mathletters}

\marge where \mbox{$a_{\pmoins} =1\pm\tf{\sqrt{2}}{2}$},
\mbox{$b_{\pmoins}= 1\pm\sqrt{2}$} and $\eps_c={\omega_c}^2/{\omega_0}^2$.  
Equations (\ref{eq:xy:ellips}) have a form identical to equations
(\ref{eq:xy}) where $\omega$ is replaced by
$a_{\pmoins}\omega_c$ and $\eps$ becomes $b_{\pmoins}\,e\,\eps_c$.
Evidently two sets of resonant frequencies occur when $\omega_c
(1+\sqrt{2}/2)$ or $\omega_c (1-\sqrt{2}/2)$ is near $\omega_0$. These
frequencies are sufficiently different that we treat them as independent
resonant solutions.  It follows that the form of the growing mode
late-time solution for $x(t)$ and $y(t)$ in an elliptical orbit is

\begin{mathletters}
\begin{eqnarray}
x(t)    & = &   \alpha'\,
                \exp\pac{\f{\sqrt{\pa{a_{\pmoins}\omega_c}^2 - 
			4\,{\omega_1}^2}}{2}\,
			\abs{b_{\pmoins}}\, e\,\eps_c t}
                \cos\pa{a_{\pmoins}\omega_c t + \theta_{\pmoins}}	\\
y(t)    & = &   \alpha'\,
                \exp\pac{\f{\sqrt{\pa{a_{\pmoins}\omega_c}^2 - 	
			4\,{\omega_1}^2}}{2}\,
			\abs{b_{\pmoins}}\, e\,\eps_c t}
                \sin\pa{a_{\pmoins}\omega_c t + \theta_{\pmoins}} 
							\label{eq:xyellips}
\end{eqnarray}
\end{mathletters}

\marge where $\omega_1$ has been defined by \mbox{$\omega_0 =
a_{\pmoins}\omega_c + b_{\pmoins}\, e\, \eps_c\, \omega_1$}. 
Here \mbox{$\tan{\theta_{\plus}} = -H_{\plus}$} and 
\mbox{$\tan{\theta_{\moins}} = 1/H_{\moins}$} where $H_{\pmoins}$ has the 
form of equation~(\ref{def:H}) with $\omega$ replaced by 
$a_{\pmoins}\omega_c$.
 Comparing to the circular orbit solution we see that the expansion
parameter $\eps$ becomes \mbox{$\f{{\omega_c}^2}{{\omega_0}^2}\,e\,
b_{\pmoins}$} so that instability occurs when
\mbox{$\abs{\omega_0-a_{\pmoins}\omega_c} < \omega_0\,
\f{{\omega_c}^2}{{\omega_0}^2}\, e\,\abs{b_{\pmoins}} /2$}.  This
expression is comparable to the circular orbit solution~(\ref{eq:width})
but resonance is ``easier'' to reach even as the strength of the tidal
force compared to the binding force (which is given by
\mbox{$\eps_c=\tf{k_c}{{\omega_0}^2}=\patf{\omega_c}{\omega_0}^2\sim
\tf{1}{{a_{\pmoins}}^2}$}) can be weaker by a factor of 3 or more compared
to the circular orbit.

Expanding $\ph(t)$ and $\eps(t)$ to order $3$ in $e$ yields additional
resonant behavior. Analogous to the above expansion we find two new
sets of excitation frequencies. The additional terms in the
differential equation for $x(t)$ and $y(t)$ above
have the form
$d_{2\pmoins}e^2\eps_c\cos\pa{2b_{\pmoins}\omega_ct}$
and $d_{3\pmoins}e^3\eps_c\cos\pa{2c_{\pmoins}\omega_ct}$ -- with 
corresponding
$\sin\pa{\cdot}$ terms as in (\ref{eq:xy:ellips}). Here
$c_{\pmoins}     = 1 \pm \f{3}{2}\,\sqrt{2}$ and            
$d_{n\pmoins}$, $n\in\{2,3\}$,  are numerical factors of order unity.
Thus, the higher order terms lead to excitations at frequencies where
$\omega_0$ is $b_{\pmoins}\,\omega_c$ (order 2) and at
$c_{\pmoins}\,\omega_c$ (order 3). The growth rates of these modes and
instability range for $\omega_0$ are smaller by corresponding powers of
$e$ so we expect to excite higher order modes only as the orbital
eccentricity increases.

\subsubsection{Projection on the line of sight}
\label{sect:projection}

An interesting, and perhaps observable signature from the simulations, is
the elliptical shape caused by the strong geometric asymmetry in the
growing mode solution for the distance of a star away from the center of
the dS. Here we consider how the induced elliptical shape varies with
respect to the line-of-sight from the MW center-of-force toward the dS
center. Consider, for the sake of generality, that the excitation is
encountered at a frequency $a\, \omega_c$ where $a$ can be chosen from the
set $\{1,a_{\pmoins},b_{\pmoins},c_{\pmoins}\}$, with the corresponding
exponential term $E_a(t)$. Thanks to equations (\ref{def:hat:r}) and
(\ref{def:hat:phi}), we obtain (where $t=0$ corresponds to perigalacticon)

\begin{mathletters}
\begin{eqnarray}
\pa{x\, \widehat{x} + y\, \widehat{y}}\cdot\widehat{r}    &=&
                E_a(t)\, \cos\pa{\pa{a-1}\omega_c\, t \label{turning:bar}
                                + \theta\vphantom{\dot{F}}}     \\
\pa{x\, \widehat{x} + y\, \widehat{y}}\cdot\widehat{\ph}  &=&
                 E_a(t)\, \sin\pa{\pa{a-1}\omega_c\, t
                                + \theta\vphantom{\dot{F}}}
\end{eqnarray}
\end{mathletters}

	In the circular case ($a=1$) the \dS{} is expanded along an
axis rotated away from the line-of-sight to the MW by an angle
$\theta$ with \mbox{$\tan\theta = -H$}.  For elliptical orbits the bar 
appears to rotate.  At order
$n$ the bar formed by the tidal interaction will be turning at a
frequency $n\,\tf{\kappa}{2}$. Thus as the \dS{} orbits between
successive perigalacticon the bar will turn $\f{n}{2}$ times.

%% file: numeric.tex
\section{Numerical Calculations and Experiments}

These analytic calculations suggest that a broad range of dS-MW orbit dynamic 
conditions can lead to parametric tidal dS excitation. 
Evidently this excitation can occur even when the orbital and dS galaxy
resonant frequencies are widely ``mistuned.'' Although our analytic
calculations simplified the force equation terms from the exact spatial
and temporal dependence
of the dS self-gravity, we can show by numerical methods how many of the
properties are retained in more realistic
self-gravitating stellar systems. 

In the discussion below we assume a system of units where the gravitational
constant $G=1$. We fix one numeric length unit 
to a physical scale of $1\U{kpc}$ and we've taken the dS to have a
total initial mass of $10^6$ solar masses. With the gravitational
constant set to unity, this implies a numerical time unit of $415\U{Myr}$.
The logarithmic external Milky Way potential has been chosen to have the form 
\mbox{$\phi (r)={v_c}^2\log(r/r_0)$}
where we assume a circular velocity $v_c=220\U{km.s^{-1}}$
(or 100 numeric units). 

We first consider the problem of a more realistic non-harmonic, but
still static, dS potential/density model. This is addressed by numerical
integration of the modified two-dimensional Mathieu differential equation.
Finally we consider direct N-body calculations to explore the collective
and time-dependent effects of the dS potential as it responds to an external
tide.

\subsection{Plummer Potential Solutions}

The Plummer model, $\phi (r)=-M/\sqrt{r^2+b^2}$ (\cf Binney and Tremaine
1987), is a rough but useful approximation for a self-gravitating dS
stellar distribution. We use this model both to initialize our numerical
simulations and in the differential equation integrations below. For
example, in this case the local cartesian $x$ acceleration of a star can
be written as $\ddot{x}=-x\,{\omega_0}^2 {\f{b^3}{{(b^2+r^2)}^{3/2}}}$
where ${\omega_0}^2=4\pi\rho_0/3$ and $\rho_0$ is the central density.
Notice that the parameter $b$ conveniently describes potentials which
range from completely unbound ($b=0$), to homogenous, purely harmonic
($b\rightarrow \infty$) models.  Substituting for the harmonic dS force
term in eqs. (\ref{eq:xy}) we obtain

\begin{mathletters}\label{eq:plumxys}
\begin{eqnarray}
\ddot{x} + x\, {\omega_0}^2 {\f{b^3}{{(b^2+r^2)}^{3/2}}} - x \,\eps\, 
{\omega_0}^2 \cos{2\ph}		\label{Mat:gen2:x}
		& = &	y \,\eps\, {\omega_0}^2\,  \sin{2\ph}	\\
\ddot{y} + y\, {\omega_0}^2 {\f{b^3}{{(b^2+r^2)}^{3/2}}} + y \,\eps\, 
{\omega_0}^2 \cos{2\ph}		\label{Mat:gen2:y}
		& = &	x \,\eps\, {\omega_0}^2\,  \sin{2\ph} \label{eq:plumxy}
\end{eqnarray}
\end{mathletters}

We were unable to construct analytic solutions to equations
(\ref{eq:plumxys}), except for the cases \mbox{$b=0$}, or
\mbox{$b\rightarrow\infty$}. Notice that this more general equation,
preserves the form of the two-dimensional parametric oscillator but with a
spatially variable ``resonance'' frequency ${\omega_p}^2(r)={\omega_0}^2\,
b^3/{(b^2+r^2)}^{3/2}$ and with
$\eps\rightarrow\eps\,{\omega_0}^2/{\omega_p}^2$. For example, stars on
circular orbits with initially constant $r$ should be tidally
excited at the first Mathieu equation resonance condition when the dS
orbit satisfies $\omega =\omega_p(r)$. We can expect that, to the extent that
individual stellar orbits aren't circular and there is a broad range in
$\omega_p(r)$, the dS tidal response should be a broad function of its
orbital frequency, $\omega$. For finite $b$ we also note that $\omega_p(r)
< \omega_0$ so that resonant behavior must occur at lower frequencies than
the dS harmonic oscillation frequency.

We can explore this solution space numerically. To distinguish stable and
unstable solutions as a function of orbital frequency $\omega$ we
numerically integrate eqs.~(\ref{eq:plumxys})  using a fourth-order
Runge-Kutta algorithm. For example, according to the analytic solution,
with $b\rightarrow\infty$, $\omega_0=4$, and $\eps =0.1$ we should obtain
solutions for $r(t)=\sqrt{x^2+y^2}$ which are growing only for dS orbits
with circular frequencies $\omega\approx\omega_0$. To verify this we
integrate an ensemble of solutions with $\omega$ between 0 and 8 over
$t=0$ to $50$ and compute the mean square radius over time for each
solution. Here $x(0)=1$ and the stellar orbits were chosen with a zero
initial velocity condition to yield radial orbits. Figure~\ref{fig:rkzero}
plots this mean solution radius for each dS orbital frequency.

Figure~\ref{fig:rkzero} illustrates several features of the analytic
parametric solution. First, instability occurs when the orbital frequency
is close to the dS resonant frequency ($\omega_p$) and, unlike in the one-dimensional parametric
oscillator, the next resonance where the orbital frequency is one-half
of the oscillation frequency is stable. This agrees with our analytic
result. It is perhaps surprising that a circular orbit with half
the orbital frequency of the dS resonance does not lead to instability --
since the tide force has 180-degree symmetry. This is a consequence of
the compressive and expansive parts of the tensor tide interaction.

The instability frequency width is determined by the tidal strength and
form of the MW potential (through $\eps$ in eq.~(\ref{eq:width})). With
$\omega_0=\omega_c=4$ and $\eps =0.1$ we expect a resonance width of 0.4
-- which is confirmed in figure~\ref{fig:rkzero}. The growth rate of
$r(t)$, as determined by eq. (\ref{eq:xygen}), is also reproduced in the
numerical calculation. For $\omega_1=0$ we expect a stellar orbit to
expand like $\exp{\omega_0\eps t/2.}=\exp{0.2t}$ which by $t=50$ yields
the amplitude plotted here.

The other extreme, where $b=0$, corresponds to a dS which is completely
unbound, i.e. where the dS self-gravity is unimportant. For example, at
late times a dS might be tidally distended so that its self-gravity
becomes negligible. A naive interpretation of the MW effect here might be
to conclude that unbound dS member stars should be tidally accelerated to
large distances from the dS center-of-mass. Figure~\ref{fig:rkinf} shows
the logarithm of the rms stellar orbit radius for the unbound ($b=0$) case
with the same initial conditions, range of dS-MW orbital frequencies, and
integration parameters as in fig.~\ref{fig:rkzero}. We see that for static
and slowly rotating tides the stellar system does indeed rapidly blow up
-- saturating our numerical dynamic range before $t=50$ for orbits near
$\omega =0$. Analytic solutions using computer aided symbolic manipulation
imply that the leading growing mode term in the solution varies as
$\exp\pac{\sqrt{\eps{\omega_0}^2-\omega^2}\,t}$. Interestingly, as this
shows, for dS orbits with short enough periods satisfying %
\begin{equation} \omega~>~\sqrt{\eps}\,
\omega_0~=~\sqrt{k}\label{eq:wcrit} \end{equation} 
localized near the dS center-of-mass, despite being unbound to the dS.

Intermediate potential models ($0 < b < \infty$)  yield unstable orbits
depending on the initial conditions of the stellar orbit.  The dS orbital
frequency which leads to instability of a dS star depends on the initial
radius of the star with respect to the dS center-of-mass in units of $b$,
and the dS central density which determines $\omega_0$. The previous two
cases have shown that dS stars which begin their orbits at normalized
distances of 0 and $\infty$ develop secular instability for dS-MW orbits
that satisfy, respectively, $\omega \approx\omega_0$ and $\omega\approx
0$. Intermediate cases lead to intermediate resonant frequencies.

The top panel of figure~\ref{fig:rk_1} plots the orbital expansion
of a star starting from $r/b = 1$ using the same integration parameters
($\omega_0=4$) as the previous Runge-Kutta solutions. Secular growth now
occurs at lower frequencies near $\omega\approx 2.8$. In a Plummer
potential-density model the half-mass radius occurs at $1.4b$. The lower panel
of figure~\ref{fig:rk1_4} demonstrates for this radius how the dS resonant
orbital frequency continues to decrease for stars at larger normalized dS
distances. Evidently stars near the half-mass radius are tidally excited
when $\omega = 1.1$. The initial 
position of a star within the dS potential helps to determine its fate in
the time-dependent MW tide.

These solutions describe a system where the tide amplitude $\eps$
and the orbital frequency $\omega$ are independent. A logarithmic MW
potential satisfies $\eps\, {\omega_0}^2 = \omega^2$. Thus for 
increasing
orbital frequency, $\eps$ in eqs.~(\ref{eq:plumxys}) is also increasing.
Qualitatively we find that this increases the orbital frequency domain over
which resonance expands the dS. Systems which are resonant with $\omega\approx\omega_p$ for fixed $\eps$
in the above discussion are now tidally disrupted for $\omega\geq\omega_p$. 

\subsubsection{Elliptical Orbits} 

Elliptical orbits are also described by eqs.~(\ref{eq:plumxys}) by taking
$\eps\rightarrow\eps(t)$ and using the actual form of $\ph(t)$ and
$\eps(t)$ for an elliptical orbit. Here we assume a logarithmic form for
the MW potential so that $\eps$ and $\omega$ are not independent. 

\begin{mathletters}
\begin{eqnarray}
\ph (t) &=& 1+e\, \sqrt{2}\sin{\kappa t} \\[3mm]
\eps(t) &=& \f{\omega^2}{{\omega_0}^2}\, \f{1}{(1-e \cos{\kappa t)^2}}
\end{eqnarray}
\end{mathletters}

\noindent where $\kappa = \sqrt{2}\, \omega$ is the epicyclic frequency.

We compute the mean distance of a star from the dS center as we did in the
previous section.  We assume a constant perigalacticon distance with
$e=1-\tf{\omega}{2}$, take $\omega_0=6.7$, $x(0)/b=1.4$, and let $\omega$
vary from 1.0 to 2.0 so that we can compare with the results of
section~\ref{elliptical:launch}. Figure~\ref{fig:ellrk} shows that a broad
range of circular frequencies lead to parametric excitation and, as we
will see, this general behavior also appears in more realistic
self-gravitating systems.  We also note that a circular orbit ($\omega=2$)
can be less dispersive than an elliptical orbit ($\omega < 2$) even if the
tide is always stronger in the circular case.

%

\subsection{Direct N-body Calculations}

In order to account for the self-gravity and 
collective dynamics of many stars within
the dS we have used a direct N-body calculation.
The numerical simulations were computed using the modified {\rm treecode
v1.4} (Barnes 1990). Typically between 1000 and 10,000 particles
were introduced into simulations which included several different
``external'' forces designed to study variously dS internal
oscillations and the tidal coupling of the stellar system to an imposed
large-scale logarithmic external MW potential.

The effects of numerical viscosity and small spatial scale potential
fluctuations can be important in these calculations and we used a range of
integration time step and potential ``softening'' parameter to understand
this dependence. Most of the simulations were done with a time step of
0.008 units. We explored the effects of the interparticle potential
softening parameter over values ranging from 0.025 to 0.3. Significant
damping of oscillations appears to be minimal in these models
with softening parameter $a > 0.1$.

With a crossing time of about 0.2 time units the relaxation timescale for
$N=1000$ simulations was at least 2 time units and much larger for softer
potentials. Many of our results only depend on measuring differential
changes between simulations, e.g. to determine the frequency
dependence in the particle ejection rate between models with different orbital
frequencies. Some results are computed from relatively long model simulation
runs. 

Several different initial particle configurations were generated
although most of the results we present here used the {\rm treecode}
Plummer model realization (\cf Aarseth et al. 1974 as implemented
by Barnes 1990).  The rms particle
radius was typically $0.1\U{kpc}$ and the Plummer parameter
was typically $b=0.16$ for a softening parameter of $a=0.1$.
To achieve this an initially isotropic
velocity dispersion of about $1\U{km.s^{-1}}$ was used and allowed
to relax before imposing any external time-dependent
tidal forces. 

We have not tried to reproduce any of the dS in detail since their MW
orbital uncertainties and the non-uniqueness of the model solutions make
detailed comparisons difficult to interpret.  Nevertheless we believe
these N-body calculations are most representative of a dS like Draco or
Ursa Minor.

\subsubsection{Oscillations}

The parametric galaxy model described by eqs. (\ref{eq:xy}) neglects 
the
effect of the oscillating potential of the dS. Thus, we can expect
normal mode galaxy oscillations to dominate even the growing parametric
tidal modes if the system's normal mode frequencies are ever excited
by the MW orbit. To explore this numerically we need to understand the
normal mode oscillation spectrum of our dS system.

Based on previous experience with particle-mesh calculations (KM)
involving $10^5$ particles or more we expected free galaxy oscillations
to be readily generated by small deviations from equilibrium
in the initial conditions of the particle configuration. For example, 
Miller and Smith (1999) had difficulty damping and suppressing
such oscillations. Even though the total energy in our
simulations of isolated dS was conserved to within 0.03 \%, we initially 
had
difficulty detecting oscillations. To excite an oscillation
spectrum we applied an impulsive radial stretching
acceleration at \mbox{$t=0$} and looked for the resulting ringing.
From a temporal fourier analysis of projections of the
particle motion onto the fundamental eigenmode, 
\mbox{$\vect{\xi} (r)=\vect{r}$},
we measure the oscillation spectrum.

In detail we compute a timeseries by summing over particles (labeled with $i$)

\begin{equation}
c_j~=~\f{1}{N}\,\sum_i\vect{r_i}(t_j)\cdot\vect{v_i}(t_j)\label{timeseries}
\end{equation}

\marge The logarithm of the temporal fourier transforms of $c_j$ from two
different simulation runs are indicated in Fig.~\ref{fig:osc1}. For an
homogeneous sphere, the power spectrum should peak at the oscillation
frequency $\omega_0$. Increasing the softening parameter both changes the
peak frequency in the power spectrum and the peak amplitude response to
the excitation. The virial equation~(\ref{oscfreq}) yields a reasonable
estimate of the observed oscillation peak frequency. For example with
softening parameter of $a=0.1$ and using the observed N-body density
distribution we compute a frequency of $\omega_0 = 3.54$, which is
consistent with the peak in Fig.~\ref{fig:osc1} (lower panel).

The shorter timestep minimizes the integration errors and the intrinsic
numerical damping noise (Figure~\ref{fig:relax:dtime}).  Similarly by
increasing the softening parameter we can decrease the effects of
integration error from the short range particle interactions. Figure
\ref{fig:osc1} confirms that as the softening parameter increases there is
an enhancement in the fundamental mode amplitude.


It is difficult to compute the true physical damping of galaxy
oscillations but comparing our results with Miller and Smith (1999)
suggests that the direct N-body calculations have a larger numerical
damping than particle-cell approximations. We expect the dS instability
growth rates from these N-body simulations to be underestimated compared
to the analytic and Runge-Kutta solutions.

\subsubsection{Parametric resonance}

In order to demonstrate Eqs.~(\ref{eq:xygen}) for a dS in a circular
orbit, we decouple the strength of the tidal force from its time
dependence. We isolate the dependence of the tide amplitude and frequency
by fixing the amplitude while independently setting the orbital frequency
($\omega$). This is analogous to the approach KM used in their
particle-mesh calculations. It allows the frequency dependence of the
resonance to be distinguished from the effect of a growing static tide
strength as the dS galactocentric distance is decreased in order to
increase $\omega$. Evidence of this resonant behavior is illustrated in
Figure~\ref{fig:circpic} which shows a snapshot of the particle positions
for three different models corresponding to $\omega = \omega_p/2$,
$\omega_p$, and $2\omega_p$. The direction of the instantaneous tide is
indicated by the line on each plot.

Figure~\ref{fig:circres} shows the effect of changing $\omega$ while
keeping the tidal amplitude (determined by $\eps$) fixed. Here the total
number of particles ejected from the dS is plotted versus time and
frequency $\omega$. Recall that $\omega_p$ is approximately 2 numerical
units and the tide amplitude corresponds to $\eps=0.5$. Successive curves
diplaced upward in this plot show how the ensemble of models have ejected
more particles at later times.

Several points are illustrated by fig.~\ref{fig:circres}. The most
important is that, as expected from the Mathieu equation solutions, the
resonant behavior (measured by particle ejection from the dS) extends to
frequencies relatively far from the effective resonant frequency
$\omega_p$. According to our solution to Eq. (\ref{eq:xygen}) and for our
choice of $\eps$ we expect an instability frequency range
\mbox{$\tf{\Delta\omega}{\omega_p}=\tf{\eps}{2}=\tf{1}{4}$} which is
consistent with the numerical results. It is also notable that there is no
second order resonance at \mbox{$\omega_p/2$}, consistent with the
analytic predictions of the coupled Mathieu equation model.

We also find in the N-body solutions that for dS circular frequencies
$\omega \le 1.5$ the ejected stars migrate toward $r\rightarrow\infty$
as the dS evolves. For $\omega \ge 1.6$ the ejected stars (more
distant than 1 kpc from the core) appear to remain in a bounded region of
space, at least over the duration of these simulations.  This result is
anticipated by eq. \ref{eq:wcrit}. In these simulations $k=2$ so the
cutoff frequency should be about $\omega = 1.4$, in fair agreement with
the N-body calculation.

\subsubsection{Elliptical orbits}	
\label{elliptical:launch}

More realistic orbital calculations also demonstrate how the dynamical
tide interaction affects the internal dynamics of a dwarf spheroidal. We
adapted the numerical calculations described above to describe a self-gravitating 
ensemble of masses orbiting within a logarithmic external potential.
We use a 1024 point \dS{} galaxy 
characterized by oscillation frequency, $\omega_0\approx 7$ numeric units.
Setting this \dS{} onto a circular orbit at $50\U{kpc}$ from the
galactic center yields a stable system with 75\% of the initial stars
and with a particle loss rate of less than 50 stars over 250 time units.
 
        To generate a family of elliptical orbits we launched the \dS{} at
$50\U{kpc}$ along the $x$ axis with variable speed $v_0 \ge v_c$ along the
$y$ axis in a constant galactic potential $\phi (r) = {v_c}^2 \log (r)$.  
Thus in our units when $\omega_c = 2$ the orbit is circular and in general
the eccentricity, $e$, varies like $e=1-\omega_c/2$. Thus, the
perigalacticon of all runs is exactly $50\U{kpc}$ and the mean tidal force
on the \dS{} is largest for the circular case ($e=0$, $\omega_c=2$) and
decreases with $\omega_c$ or as the eccentricity~$e$ increases.

	Figure~\ref{fig:eccentricity} shows how the number of particles
ejected from the \dS{} varies with orbital ellipticity and time. Here the
orbit time is measured from the injection of the equilibrated \dS{} into
the logarithmic potential. The broad frequency response of the dS in an 
elliptical orbit is confirmed from the Runge-Kutta solutions 
(Figure~\ref{fig:ellrk}).
It is interesting that elliptical orbits, with
weaker tides than a circular orbit can lose more stars. Eccentric variable
tides can eject stars where a stronger constant tide cannot.

\subsection{Velocity Dispersion}

The velocity dispersion of our parametrically excited dS galaxies
varies by a large factor over their orbits and can be a strong function
of the extent of the dS over which the dispersion is calculated. For example a general
feature of these models is that the dispersion increases outward
from the core of the dS. Nevertheless it is possible
to inflate the dispersion by an order of magnitude or more depending on
the dS resonant and orbital conditions. For example, in Fig.~\ref{fig:dsveldisp}
we plot the instantaneous line-of-sight velocity dispersion for the
model above with orbital eccentricity of 0.5. While this is a relatively
high eccentricity orbit, it is not particularly close to resonance (with a
slow growth rate) and we plot the dispersion after the growing mode is established. It is clear that at late times the dispersion
increases by more than a factor of 10 yielding virial mass estimates
exaggerated by a factor of 100. Figure~\ref{fig:dsvelpic} shows this
model when the dispersion was 10 times larger than its initial 
configuration.
This system is not particularly close to dissolution and it has
lost only $300$ stars out of 734 when it was launched in its orbit. It also seems to retain its central concentration
and surface brightness profile. 

\subsection{Interpreting galaxy morphology from N-body calculations} 



Our analytic parametric oscillator calculations and numerical integrations
of more realistic Plummer model systems showed that resonant excitation of
dS stars should produce elliptical systems with their long axis rotated by
an angle $\theta$ away from the direction toward the force center in the
plane of the orbit. For a circular or elliptical orbit this angle is
calculated from $\tan\theta = -H$ (section~\ref{sect:circular}). At the
first parametric resonance this angle is 45 degrees. Further from the
resonance condition this angle varies from an orientation along the
center-of-force direction to being perpendicular to it. The existence of a
nonzero dS bar angle is evidence of parametric resonance.

We found in section~\ref{sect:projection} that this angle also describes
the dS elongation in an elliptical orbit at perigalacticon. During the
orbit the bar of a \dS{} should rotate with respect to the center-of-force
direction with increasing frequency, depending on the order of the
resonance in orbital ellipticity, $e$.

We have computed the direction of the dS bar from the moment of inertia
tensor of the dS stars. Figure~\ref{fig:snapshot} shows a series of
snapshots from the simulation with $e=0.22$. In this figure the
center-of-force is always to the left and $\theta (t)$ is the angle
between the two solid lines which intersect at the center of the dS.
Figure~\ref{fig:first:order} shows how the angle varies with time for
this calculations. Each spike in the graph corresponds to one rotation of
the bar. We find that the rotation rate increases with increasing
eccentricity (as it should) as higher order modes in powers of $e$ are
excited.

The bar rotation indicated in Fig.~\ref{fig:first:order} is not strictly
sinusoidal with the epicycle period. We also note that at perigalacticon
our dS systems tend to have the bar directed toward the force center
($\theta = 0$). Recall from our discussion of the Plummer model that
not all stars in the dS are characterized by the same resonance frequency
$\omega$, so that stars near the center of the dS have larger
frequencies than stars at larger distances.  
Thus for given orbital circular and epicyclic frequencies the response
of many of the dS stars is non-resonant. Thus near perigalacticon many of the
stars (especially near the core of the dS) simply respond to the tide like
any non-resonant fluid system. Our moment of inertia calculation is thus
controlled by contributions from both resonant and non-resonant stellar
components. This can lead to a more complex behavior of the bar
pointing direction than our model predicts in detail.

The outer parts of the dS (see Fig.~\ref{fig:snapshot}) are curved as
if $H$ depends on distance from the dS. This is anticipated from
the Plummer calculations since the
effective $\omega_p$ decreases outward. The curvature in the bar at
large distances from the dS raises the possibility that snapshot 
observations
of real tidally excited stellar systems may provide useful constraints on
the form of the dS potential.

%% file: observed_comparisons.tex
\section{Comparing the parametric oscillator model with observations}

Our earlier comparisons of dS dynamics and morphology with a simpler
resonant tidal excitation model depended on the commensurability of the dS
oscillation frequency (and damping width) and its MW orbital driving
frequency to generate elliptical systems with large velocity dispersions.  
The parametric oscillator model presented here more accurately describes
the dS-MW interaction and is a more precise explanation for dS properties.
In particular their elliptical shape and sometimes large velocity
dispersions is a natural consequence of this model.

Parametric dS galaxy oscillations effectively
increase the dS gravitational interaction cross-section with the MW.
Properly accounting for the time-dependent tidal interaction of a dS
with the MW, even when the dynamical frequency of the dS
($\omega_0$) is not commensurate with its orbital or epicyclic
frequency around the MW, is critical to describing the dS dynamics. We
have shown how the frequency domain near the dS fundamental frequency
which describes resonance is not characterized by the galaxy
oscillation damping time but by the tide amplitude
($\eps$).  For a logarithmic MW potential in a circular orbit the
fractional resonant-growth frequency domain is simply
${\omega_c}^2/{\omega_0}^2$ (the square of the ratio of the circular orbit
frequency and the internal galaxy oscillation frequency). Equation 
\ref{eq:width} implies that any
dS with oscillation frequency less than about 1.4$\omega_c$ will be
tidally disrupted.

As we noted above, elliptical orbits can be even more likely to disrupt dS
stellar systems because of their rich epicyclic harmonic structure.
Eccentric orbits lead to disruption of ``stiffer'' dS with internal
frequencies as large as 1.7, 2.4, and 3.1 (or larger) times the circular
orbital frequency -- corresponding to 1st, 2nd, and 3rd order terms in
orbital eccentricity. Note that parametric resonance increases the mean dS
central distance of its member stars and their velocities (and velocity
dispersions). The exponential growth time depends on how close the dS is
to a resonance condition but as eq.~(\ref{eq:xygen}) shows the growth
times of, for example, the velocity dispersion can be comparable to the
orbit period.  In general we expect larger growth rates as the ratio of
circular to internal dS frequency increases toward unity.

Despite the undetermined orbital characteristics of the dS, it is
interesting to compare our best estimate of $\eps$ for each of the MW
dwarf spheroidals with their kinematic properties. It is our contention
that the dS kinematics are not dominated by dark matter halos, but by
their MW tidal interaction. Thus we estimate their mass from their
luminosity and not their velocity dispersion, since the dS are not in
virial equilibrium.  Core (half-light) radii ($r_c$) and galactocentric
distances ($d$) are used from the compilation by Mateo (1998). A pulsation
frequency $\omega_0$ is computed from the VanderVoort approximation
$\omega_0 = \pi\sqrt{G\rho_0}$. We compute the central density $\rho_0$
from the observed total luminosity (assumed equal to the total dS mass in
solar units) and the expression $\rho_0=\f{3M(0.64)^3}{4\pi r_c^3}.$ This
is exact when the density distribution equals a Plummer model and is a
reasonable empirical approximation for the MW dS. We also take
$\omega_c=v_c/d$ (with $v_c=220\U{km.s^{-1}}$) as the circular frequency
around the MW.  Adequate for our purposes, we note that the parametric
growth rate increases with $\eps$ and the ratio $A=\omega_c/\omega_0$.  
If the dynamics of the dS are dominated by tidal interactions we expect
observed dS virial ``M/L'' ratios to increase with $A$.  
Figure~\ref{fig:dsprops} plots $A$ against M/L.  The general trend of
increasing ``M/L'' with $A$ is in good agreement with the parametric
model, especially given the uncertainty in the dS orbital parameters.

Our ``snapshot" knowledge of the Local Group Dwarf Spheroidals means
that we cannot measure their orbital ellipticity (although see Kuhn 1993 for 
a discussion of this point) nevertheless, knowledge
of the dS radial velocities and galactocentric distances is sufficient
to suggest that at least the nearest dS must be tidally inflated by
parametric resonance.

Another interesting and perhaps observationally verifiable consequence
of parametric excitation is that the long axis of the dS bar should, in
general, be inclined with respect to the separation vector between the
MW center and the dS. Orbits which are nearly circular should yield a
dS with the leading edge closer to the MW than the trailing edge and
inclined at an angle of 45 degrees.  Elliptical orbits produce a
rotating bar whose phase with respect to the radius vector direction
depends on how close the dS is to resonance and the proximity of the dS
in its orbit to perigalacticon.

%% file: conclusions.tex
\section{Summary and Conclusions}

We have demonstrated an analytic model of time dependent dS-MW tidal
interactions which extends the Mathieu Equation for parametric
oscillations to two dimensions. Several important general conclusions
follow from this model:
\begin{enumerate}
 \item Secular instability can cause dS star orbits and velocities to
grow, depending on the ellipticity and circular frequency of the MW orbit
and the resonant frequency of the \dS.
 \item Exponential growth can occur over a broad range of frequencies, not
just near the characteristic dS harmonic frequency. Growth times can be
comparable to the MW orbital period.
\item Direct integration of more realistic, non-harmonic dS potentials,
yield the same qualitative behavior predicted by the Mathieu equation.
\item Self-gravitating N-body calculations also confirm the existence of
broad resonance conditions that lead to stellar ejection from the dS
system, elliptical bar formation, and inflated (non-virial) velocity
dispersions.
\item Model expectations for the parametric growth rate in MW dwarf
galaxies have been contrasted with the observed  M/L measurements for
the 8 Milky Way dS.  The expected trend of increasing M/L with
$\omega_c/\omega_0$ is confirmed.
\end{enumerate}

Considerable attention has been given to finding dark matter density
distributions which will both stabilize the luminous components of the dS
so they survive MW tidal encounters, but not so much that they don't allow
stars to be ejected to populate ``tidal tails''. These tails are now
clearly observed by several methods (Kuhn et al. 1996, Smith et al. 1997,
Martinez-Delgado et al 2001) and are an observational hurdle that any dS
model must pass.  While it appears that there may be dark matter models
which cannot be ruled out (\cf Mayer et al. 2002) here we have developed a
consistent model of MW-dS interactions which does not require any
invisible dS mass component and which appears to account for the
morphology and dynamics of the dS.  The prediction of extratidal dS stars
in the simple KM resonance model were later observed. We can hope that the
considerably refined predictions of the parametric oscillator model (for
example the inclined bar) may also be empirically established.

\bigskip

{\bf Acknowledgements}

We're grateful to Josh Barnes for help using his \ttt{treecode v1.4}.

JJF is grateful to the IfA for support in many forms 
during his internship from the ENS -- 
and to his grand-father who left us while this work was being done.

%% file: PS/tide.tex
\begin{figure}
\plotone{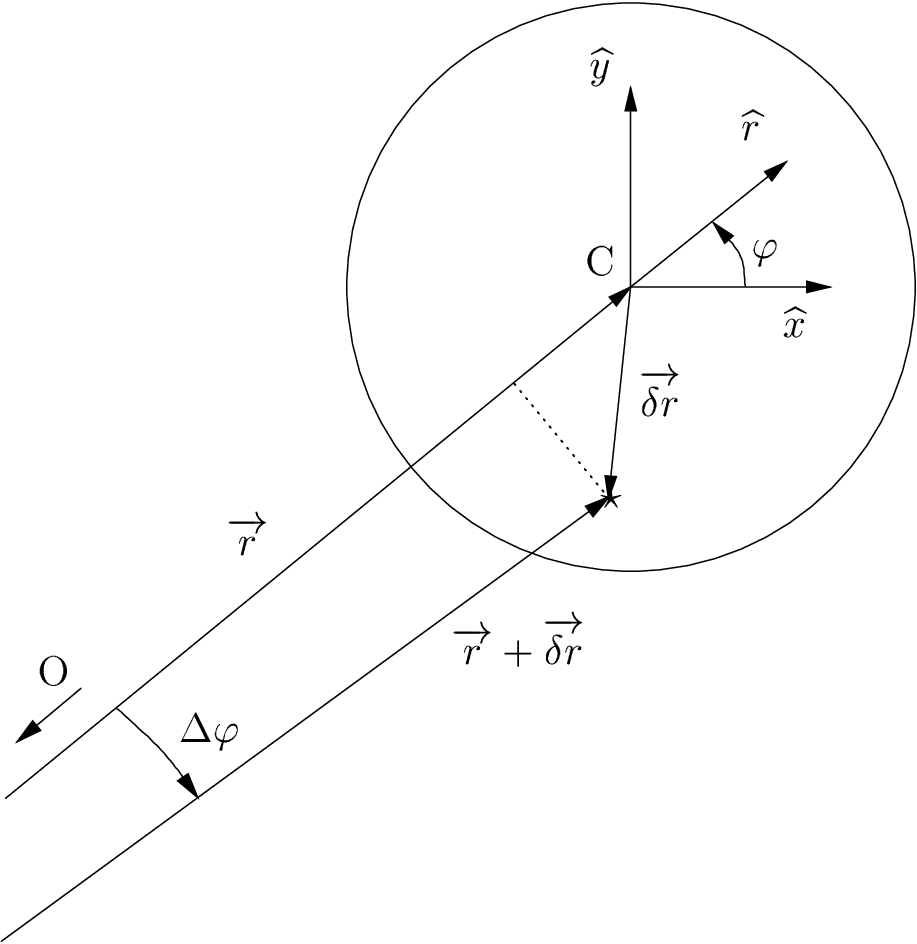}
\caption{\textbf{Tidal interaction geometry}\label{fig:tide}
A dS star is represented by $\vec{\delta r}$ in the dS-frame centered on
$C$ and by $\vec{r}+\vec{\delta r}$ in the MW-frame of reference
centered on $O$.  }
\end{figure}

%% file: PS/rkzero.tex
\begin{figure}
\plotone{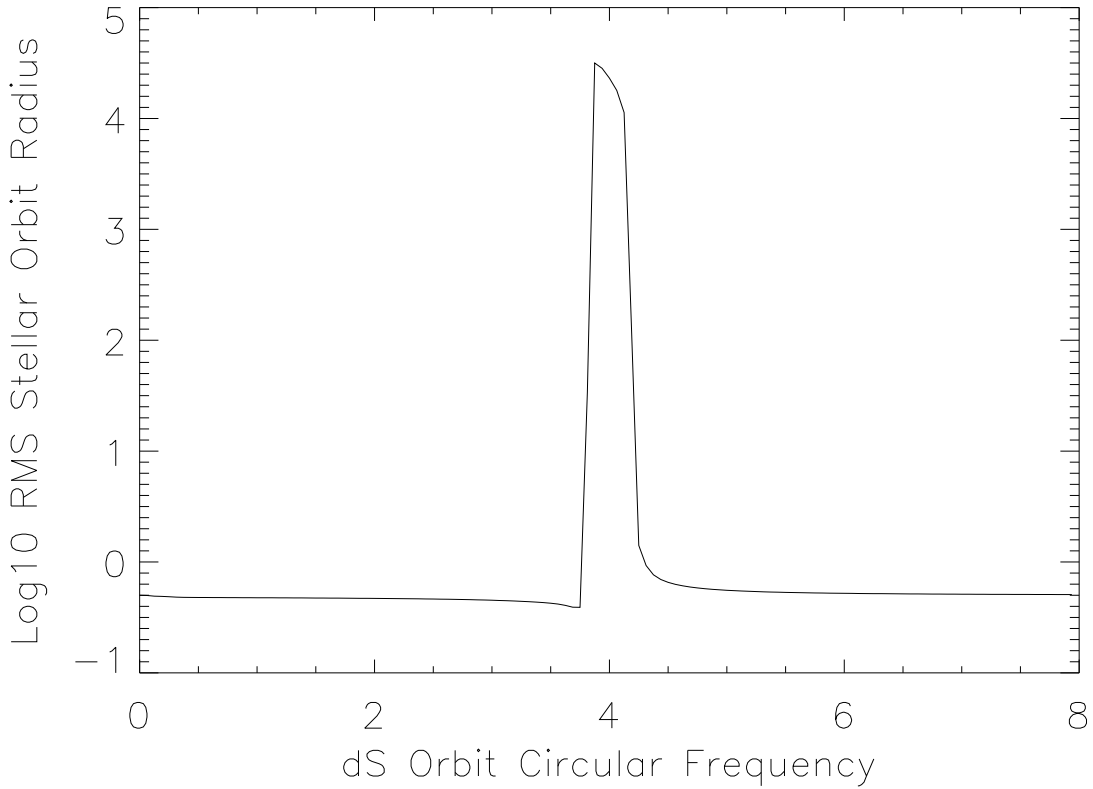}

\caption{\textbf{$b=\infty$, harmonic case \label{fig:rkzero}}. Decoupling
the tide amplitude and circular frequency allows us to see where the
parametric excitation takes place. For this Runge-Kutta integration over
50 time units, we used $\omega_0 = 4$, $b= \infty$, $\eps=0.1$ with
initial conditions $x(0)=1$ and $\vec v = 0$.  The resulting RMS stellar
orbit radius peaks for dS circular frequencies near $\omega_0$ with a
half-width consistent with $\tf{\eps\, \omega_0}{2} = 0.2$ as predicted by
the analytical solution. Note also that the one-dimensional parametric
oscillator excitations at $\tf{\omega_0}{n}$ with $n\ge 2$ do not appear,
as predicted by our analytic model.}
\end{figure}

%% file: PS/rkinf.tex
\begin{figure}
\plotone{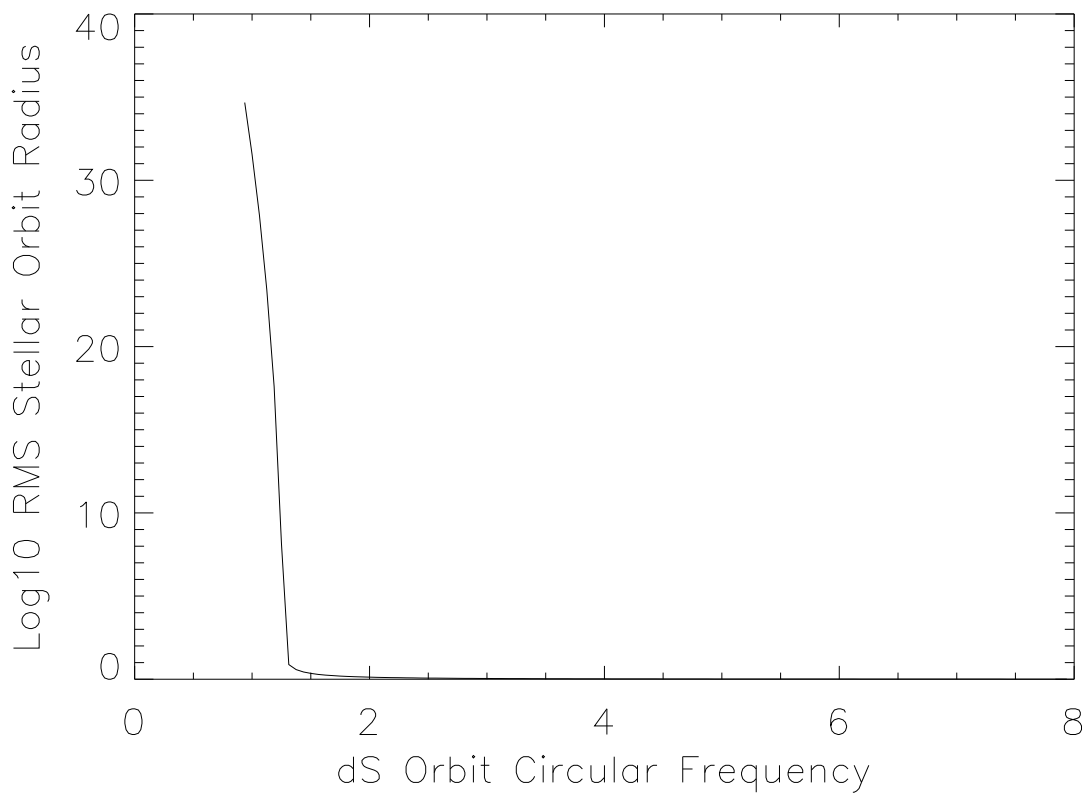}

\caption{\textbf{$b=0$, unbound case.} \label{fig:rkinf}
Symbolic computer aided analytic solutions for this unbound case 
suggest that the dominant term should be of the form 
$\exp{\sqrt{\eps\, {\omega_0}^2-\omega^2}\, t}$, which gives a cut-off 
frequency $\omega_\text{cut} = \sqrt{\eps}\, \omega_0 = 1.26$ as observed 
here. Stellar orbits are stabilized (bounded) 
when $\omega > \sqrt{\eps}\, \omega_0 = \sqrt{k}$ where $k$ is the 
tidal constant defined in eq.~(\ref{def:k}). }
\end{figure}

%% file: PS/rk_1.tex
\begin{figure}
\epsscale{.75}
\plotone{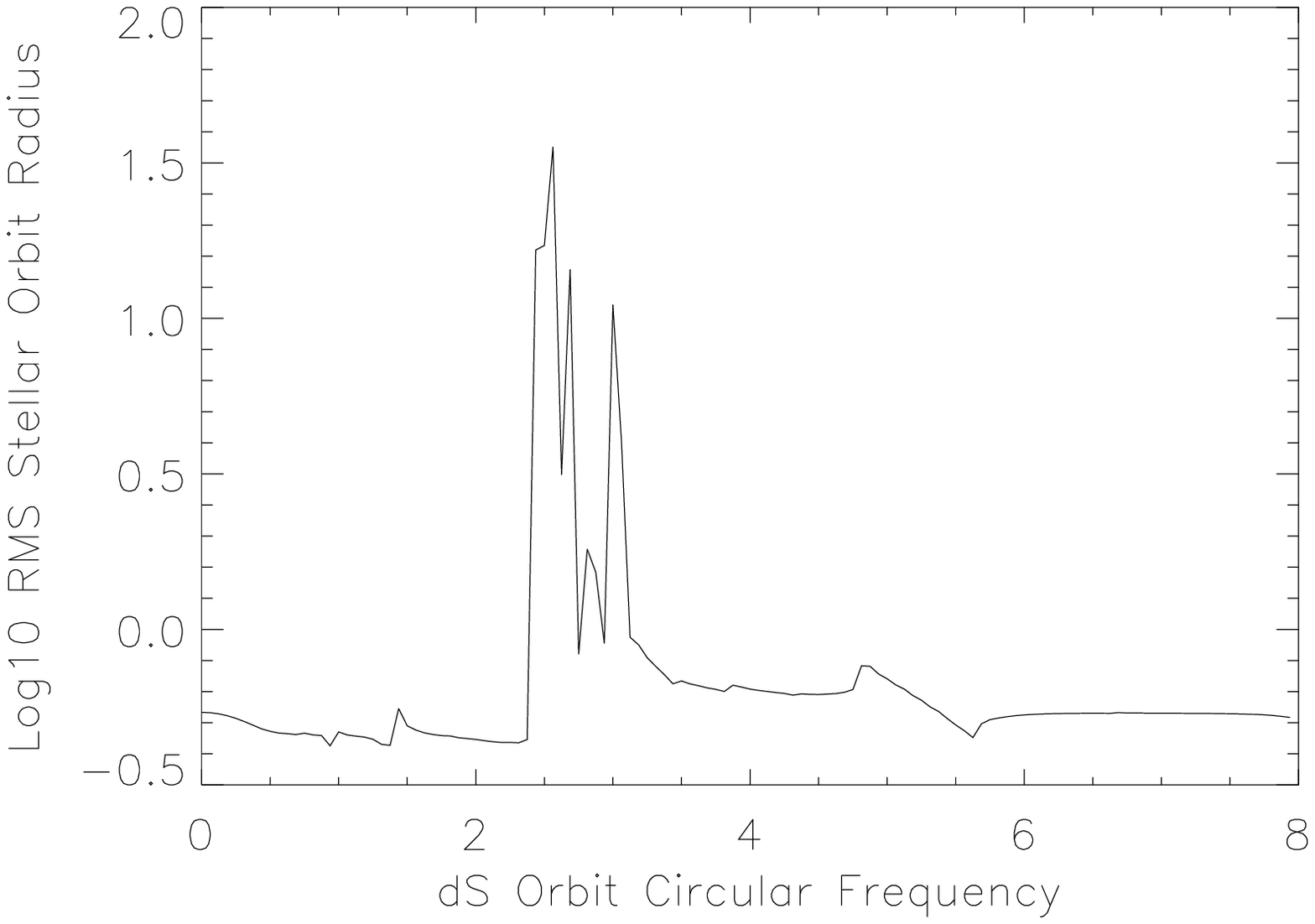}
\plotone{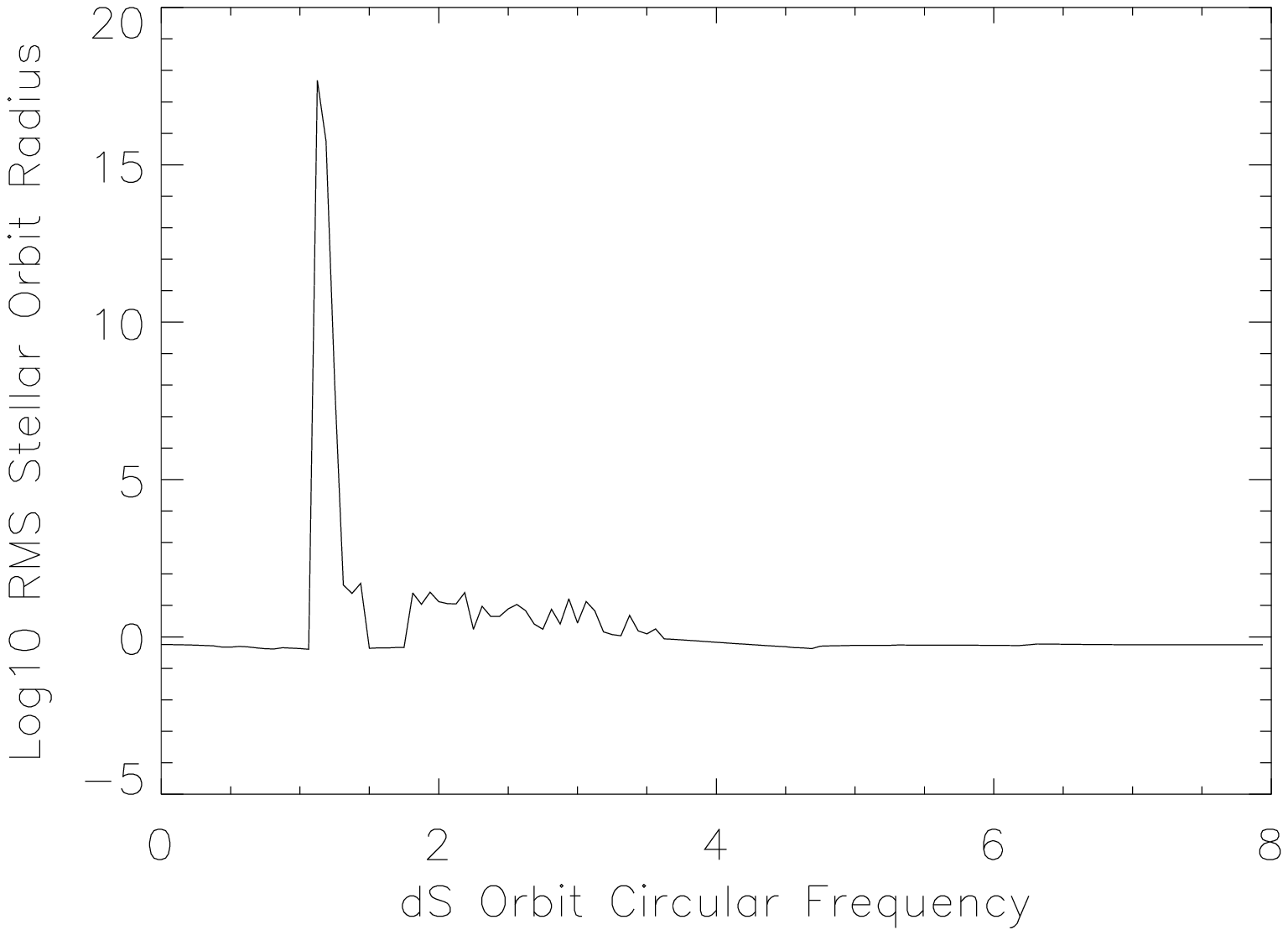}
\epsscale{1}
\caption{\textbf{$r/b=1$ (top panel) and $r/b=1.4$, (lower panel)
intermediate cases.} 
\label{fig:rk_1}\label{fig:rk1_4}
Conditions are identical to figure~\ref{fig:rkzero} except that $b$ has
now be chosen so that $\tf{x(0)}{b} = 1$ and $\tf{x(0)}{b} = 1.4$.
The apparent excitation frequency $\omega_p$
decreases as you move further away from the core.
}
\end{figure}

%% file: PS/ellrk.tex
\begin{figure}
\plotone{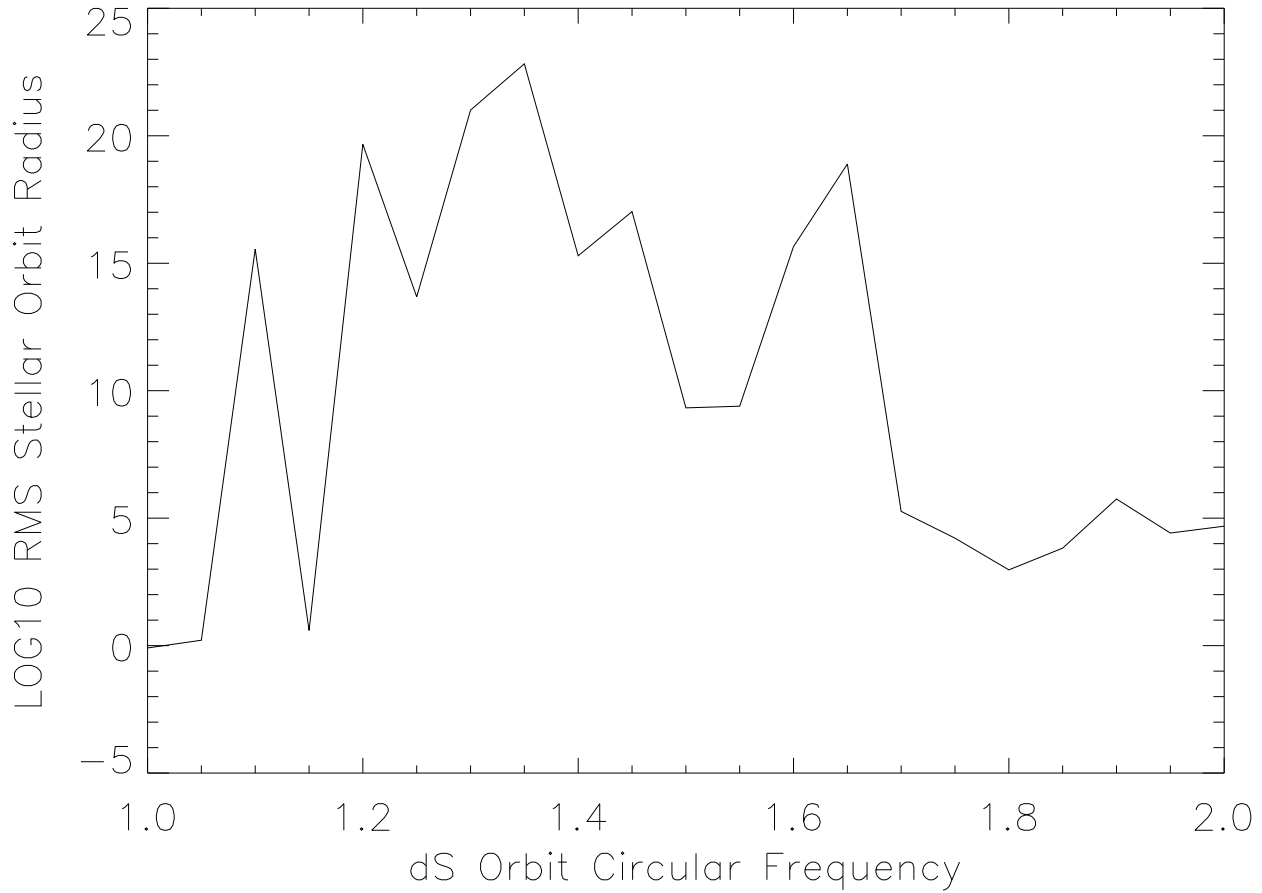}

\caption{\textbf{Elliptical orbit excitation\label{fig:ellrk}}.
Runge-Kutta integrations for a \dS{} with $\omega_0=6.7$, $x(0)/b=1.4$ in 
elliptical orbits with fixed perigalacticon distance.
Here $e = 1-\omega/2$ so that $\omega=2.0$ is a circular orbit. The tide amplitude and circular frequency are constrained as they are in a logarithmic potential so that $k = \omega^2.$
Orbital and dS parameters correspond approximately to the N-body simulation parameters. The broad frequency response of the dS is a consequence of
elliptical orbits.}
\end{figure}

%% file: PS/eps_fft_bon.tex
\begin{figure}
\epsscale{.7}
\plotone{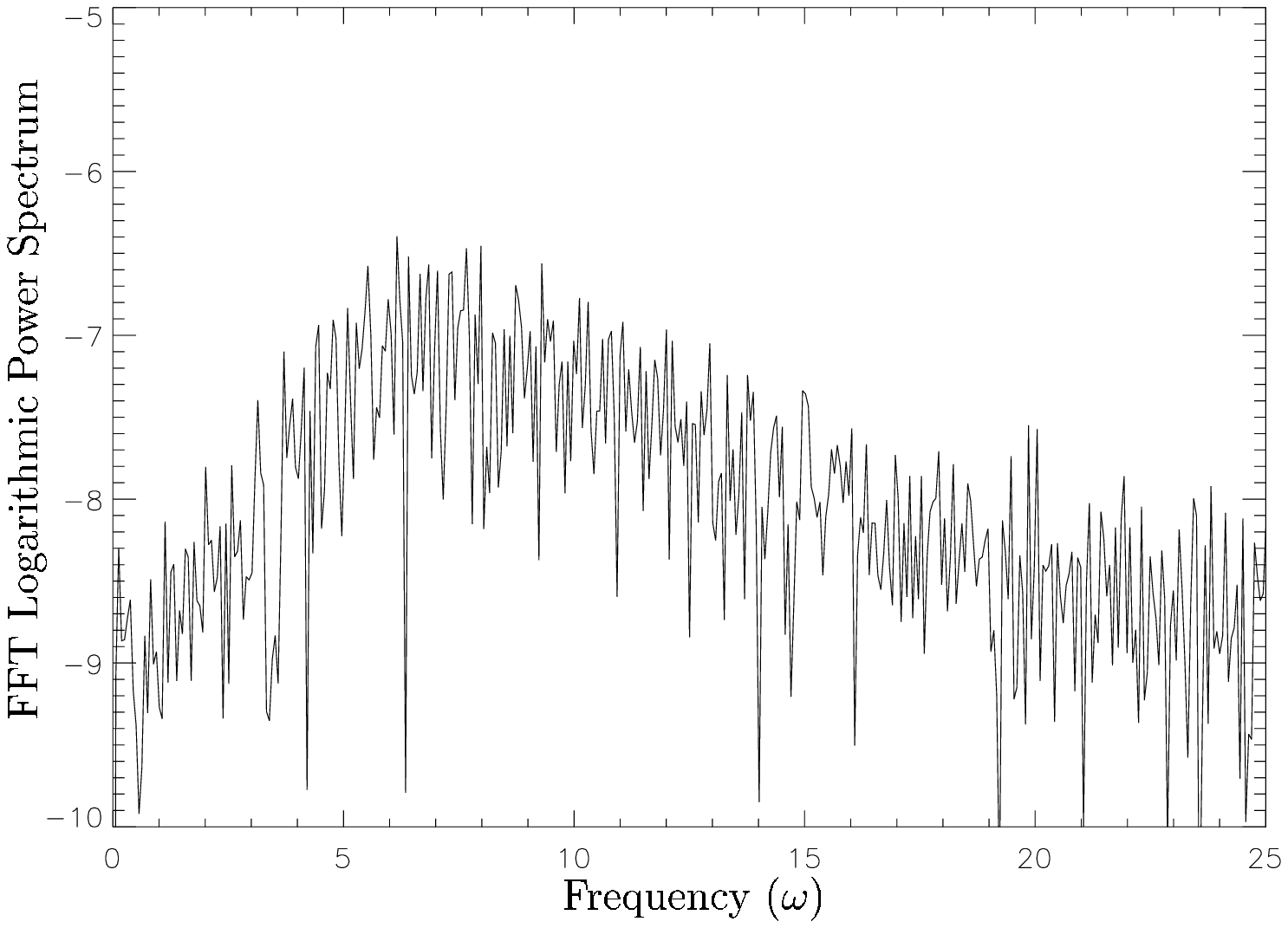}
\plotone{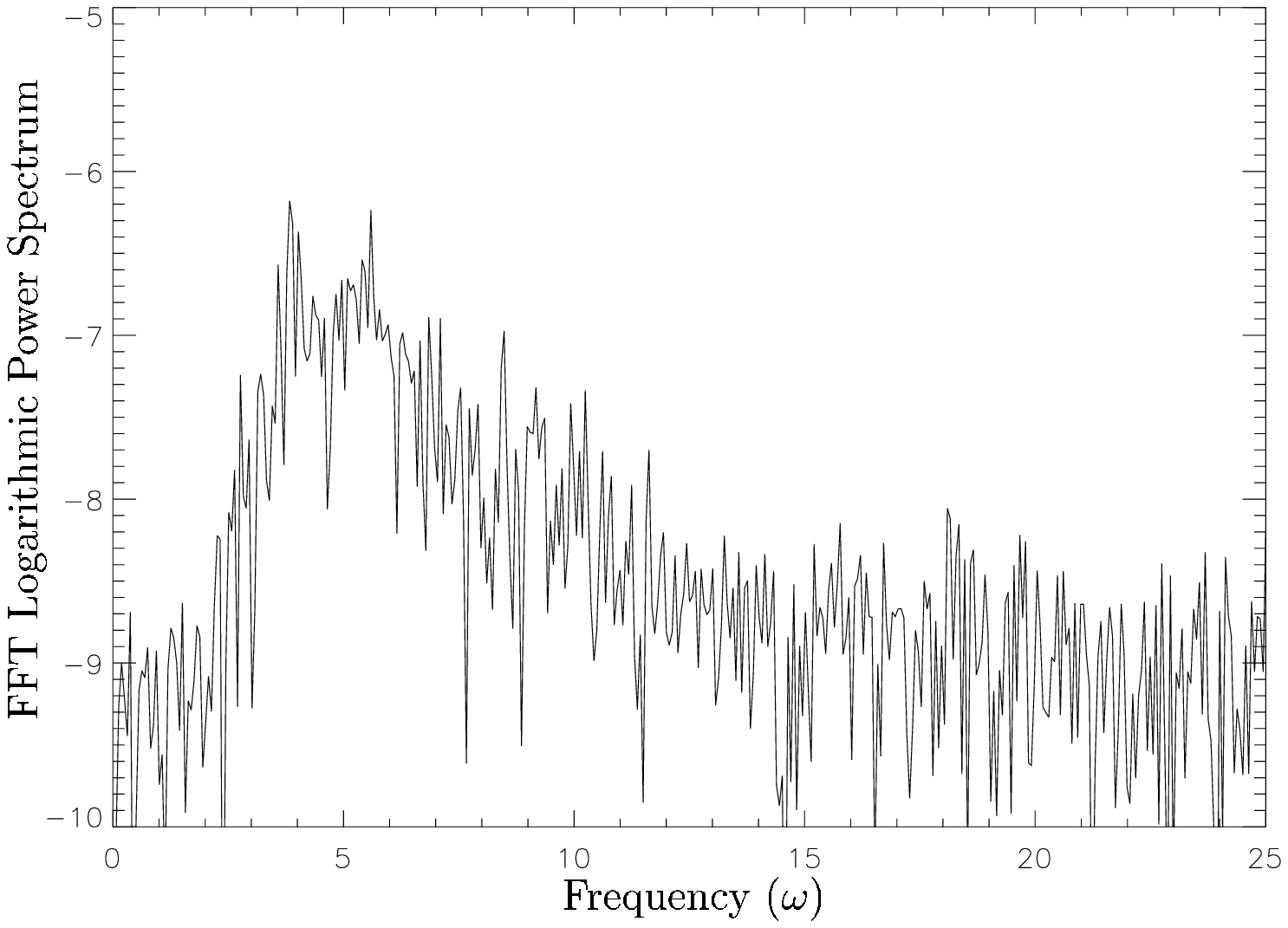}
\caption{\textbf{Influence of the softening parameter.}\label{fig:osc1}
\label{fig:fft}
 The potential softening parameter also affects the relaxation time and
numerical dissipation. Here we plot the galaxy oscillation spectrum for
identical initial N-body conditions but with softening parameter values or
$a=0.025$ (upper panel) and $a=0.1$ (lower panel).
Curves are generated from the timeseries defined in
equation~(\ref{timeseries}). }	
 %
 %
 %
\epsscale{1}
\end{figure}

%% file: PS/relax_dtime_bon.tex
\begin{figure}
\plotone{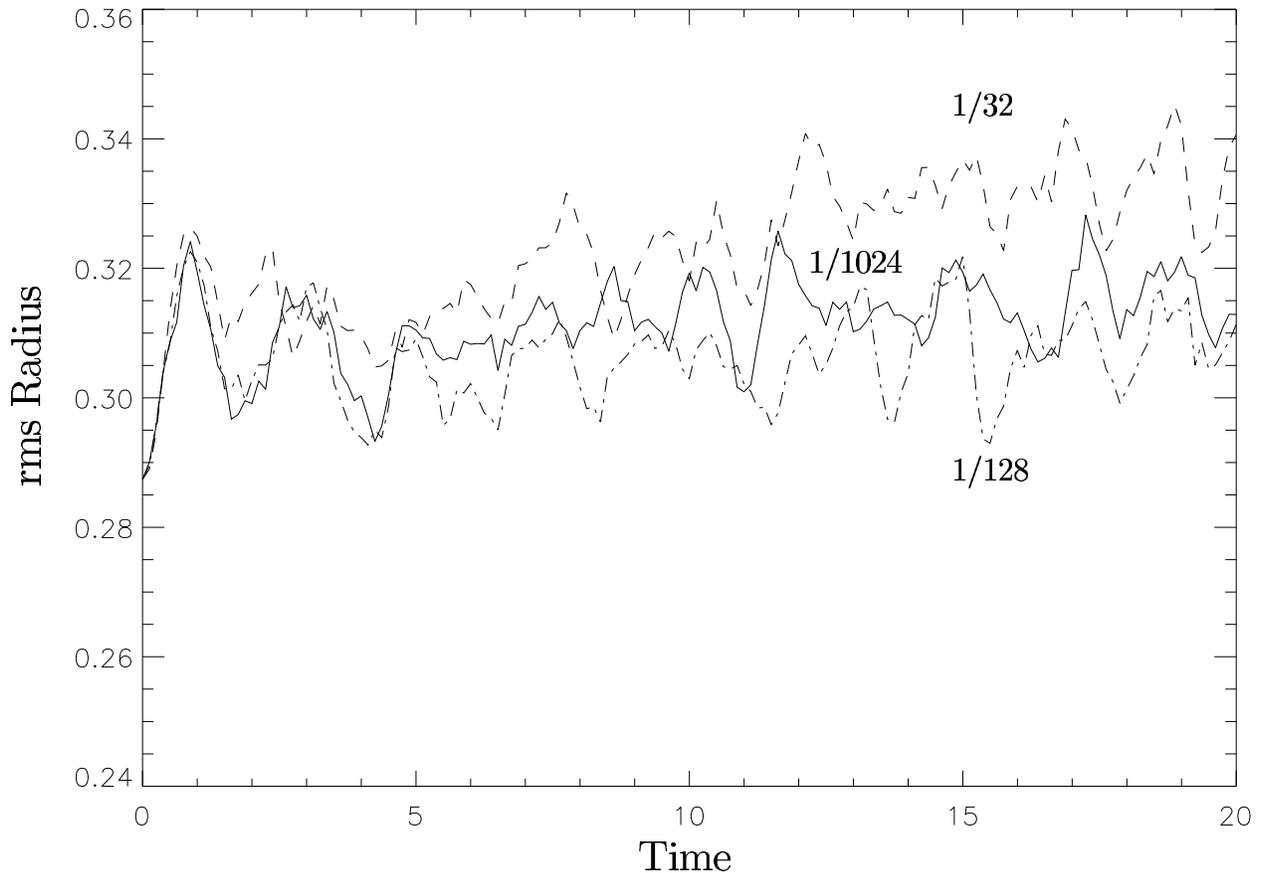}
\caption{
\textbf{Influence of integration time-step on relaxation.}
\label{fig:relax:dtime}
 To produce long simulations where numerical dissipation is minimized, the
integration time-step must be decreased. These curves show the evolution
of isolated sytem rms radius for 3 runs with integration time-steps of
$\tf{1}{32}$, $\tf{1}{128}$ and $\tf{1}{1024}$ time units. The radius of
the system grows appreciably for $\tf{1}{32}$ time-steps but is
essentially stable for the smaller integration times.  We adopt
$\tf{1}{128}$ in our integrations but note that the fine structure in
these calculations (consider the difference between the two shorter
integration rms curves) may not be reliable.}
 \end{figure}

%% file: PS/omega_snapshot_bon.tex
\begin{figure}
\epsscale{.9}
\plotone{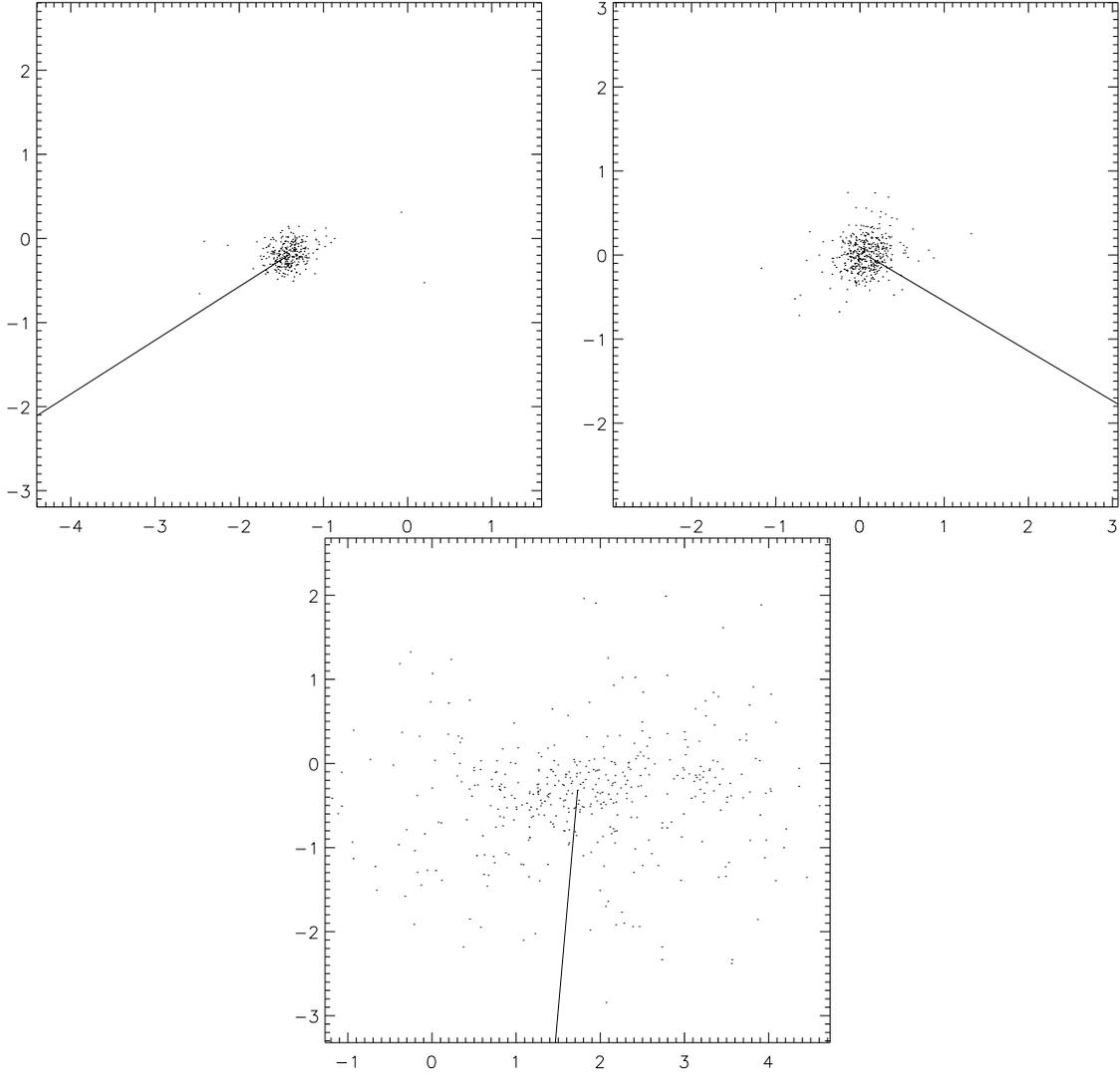}
\epsscale{1}
 \caption{\textbf{Snapshots at $\tf{\omega_p}{2}$ (upper left), 
$2\omega_p$ (upper right)
and $\omega_p$ (lower panel).}
 \label{fig:circpic}
Snapshots of the projected dS star distributions 
have been taken at $t=25$ numeric units for three runs ($\omega =
0.8$ (upper left), $\omega=4$ (upper right) and $\omega=1.7$ (lower 
panel)) 
corresponding to
$\omega\approx\tf{\omega_p}{2}$, $\omega \approx 2\omega_p$ and
$\omega\approx\omega_p$. The instantaneous direction of the
tide is indicated by the solid line. 
Snapshots at $\tf{\omega_p}{2}$ and $2\omega_p$ are very
similar: the \dS{} does not seem to be affected by the tide. The 
apparent radius is smaller for case
$\tf{\omega_p}{2}$ because almost a third of the stars have been
ejected to infinity, although even in this case, the dS shape is not
affected.
On the contrary, the snapshot at $\omega_p$ shows a \dS{} that has been
completely disrupted by the tidal interaction. Its elliptical shape
is turning with the tide, but keeping an approximately
constant angle with the line-of-sight -- consistent with the
parametric predictions for circular orbits.}
 \end{figure}

%% file: PS/omegaomega_bon.tex
\begin{figure}
\plotone{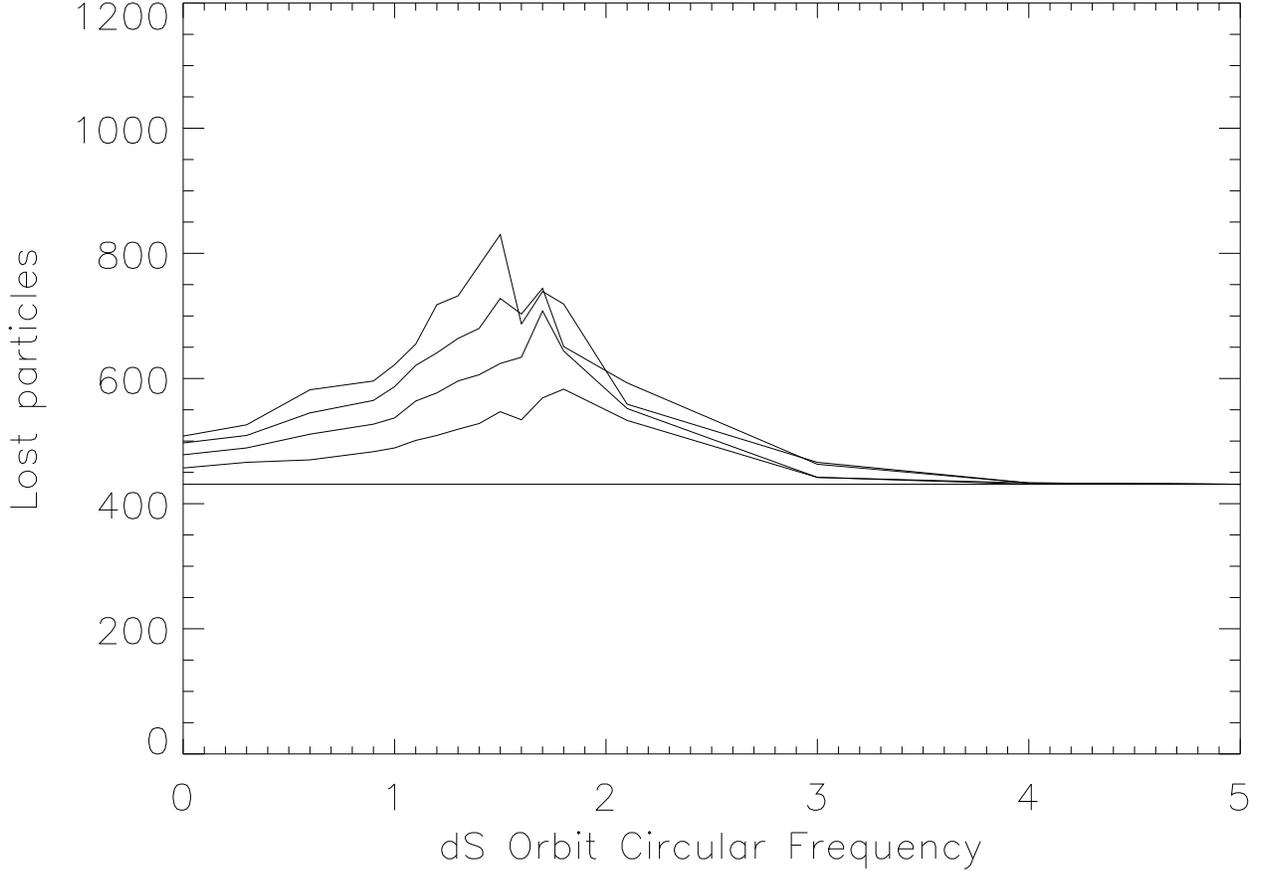}
\caption{\textbf{Number of particles lost versus circular orbit 
frequency.}\label{fig:omegaomega}\label{fig:circres}
 These simulations were performed using a simple turning tide with a
constant amplitude $k=2$ corresponding to $\eps\approx 0.5$. Particles are
defined to be lost when they reach a distance greater than $1\U{kpc}$ from
the center of mass position. The multiple curves plot the number of
particles lost at successive times. The curves are separated by 6.25 time
units. The effective resonant frequency of the dS is $\omega_p = 1.7$
initially. As stars are lost and the central density decreases, $\omega_p$
decreases and we also observe the resonant peak shift to lower
frequencies. We also observe two regimes in the simulation results -- if
$\omega\leq 1.5$, lost stars are ejected to reach $r=\infty$, but when
$\omega\geq 1.6$, stars remain in a bounded region of space near the dS.}
\end{figure}

%% file: PS/100by100_ejection_bon.tex
\begin{figure}
\plotone{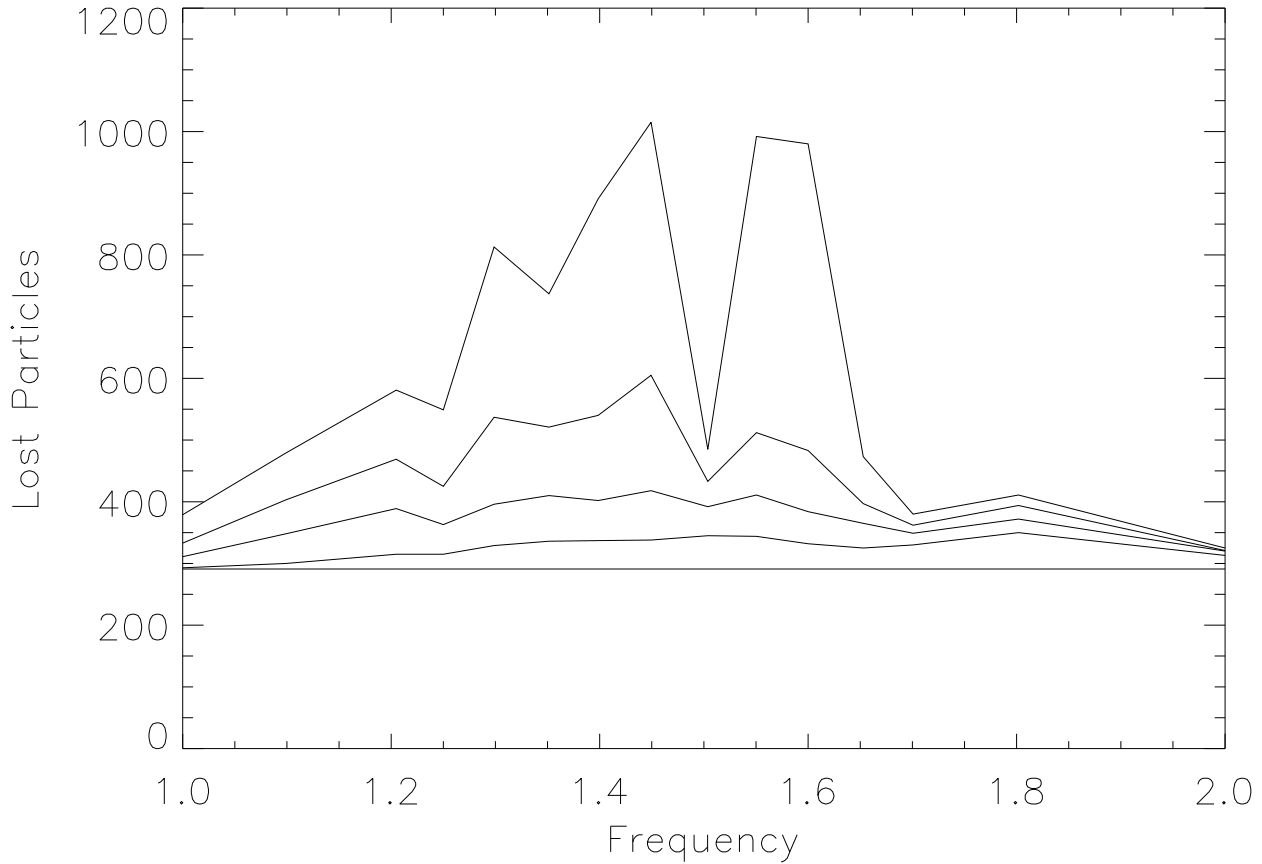}
\caption{\textbf{Parametric excitation on elliptical orbits.}
\label{fig:refined:resonance}\label{fig:eccentricity}
 This figure plots the variation in the number of particles lost versus
time and orbit frequency. Multiple curves in the figure show the dS system
at successively later times $\Delta t=25$.  With a fixed perigalacticon
distance at $50\U{kpc}$ we have $\omega_c\approx2\pa{1-e}$, so that
$\omega_c=2$ corresponds to the circular case $e=0$. The maximum loss is
1024 stars but all simulations began with 734 stars when injected into
their MW orbit.
 Elliptical orbit tidal interactions can expand and entirely disrupt a
galaxy due to parametric excitation where a stronger static tide has only
a small effect. These results compare favorably with the non-gravitating
Runge-Kutta elliptical orbit calculations above (\cf 
figure~\ref{fig:ellrk}).} 
\end{figure}

%% file: PS/veldisp.tex
\begin{figure}
\plotone{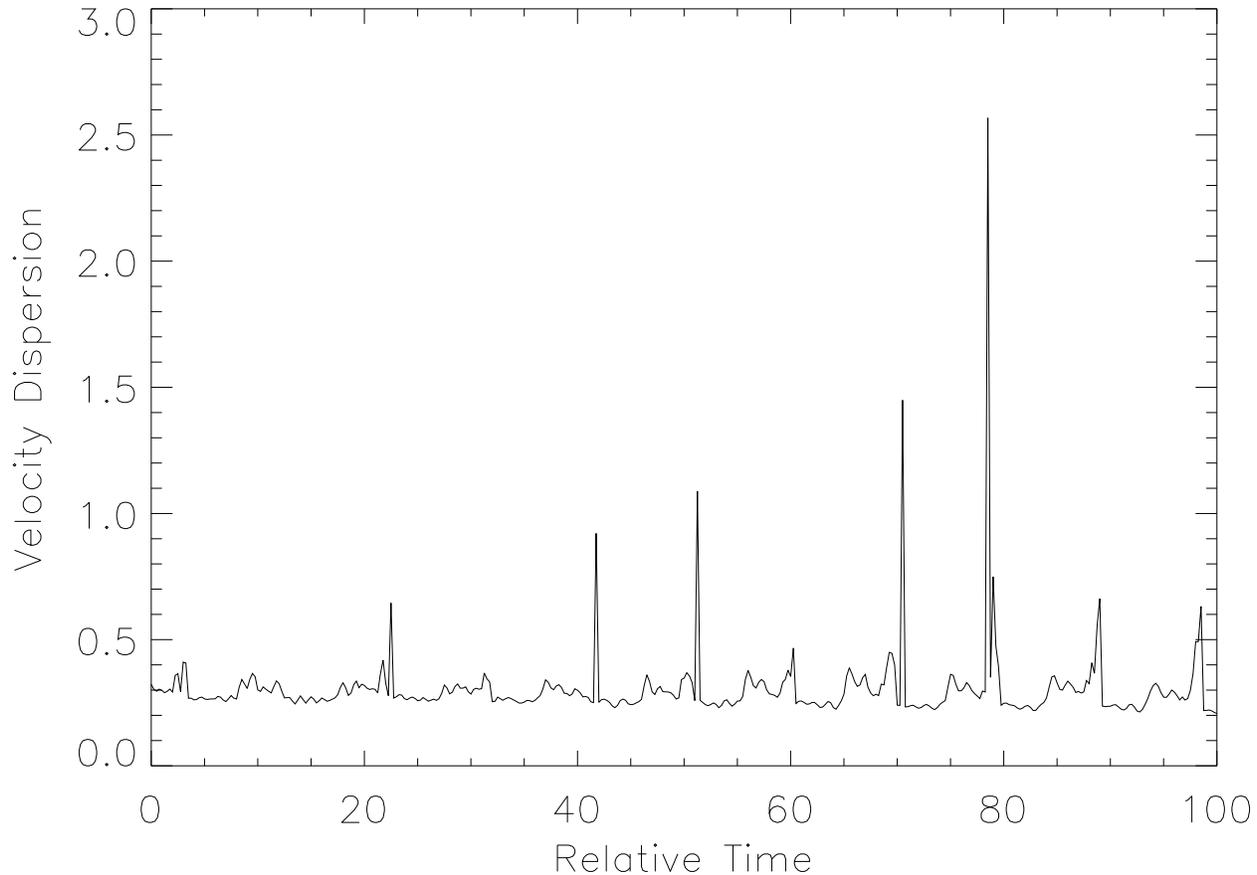}

\caption{\textbf{Velocity dispersion.}\label{fig:dsveldisp}
This figure plots the derived line-of-sight velocity dispersion
near a weakly resonant system with orbital eccentricity 0.5.
The fluctuating velocity dispersion can be many times larger than
its initial equilibrium value before the dS appears to be disrupted. 
Virial mass estimates near the perigalacticon dS passages could yield
M/L values exaggerated by a factor of 100.
}
\end{figure}

%% file: PS/dsvelpic.tex
\begin{figure}
\plotone{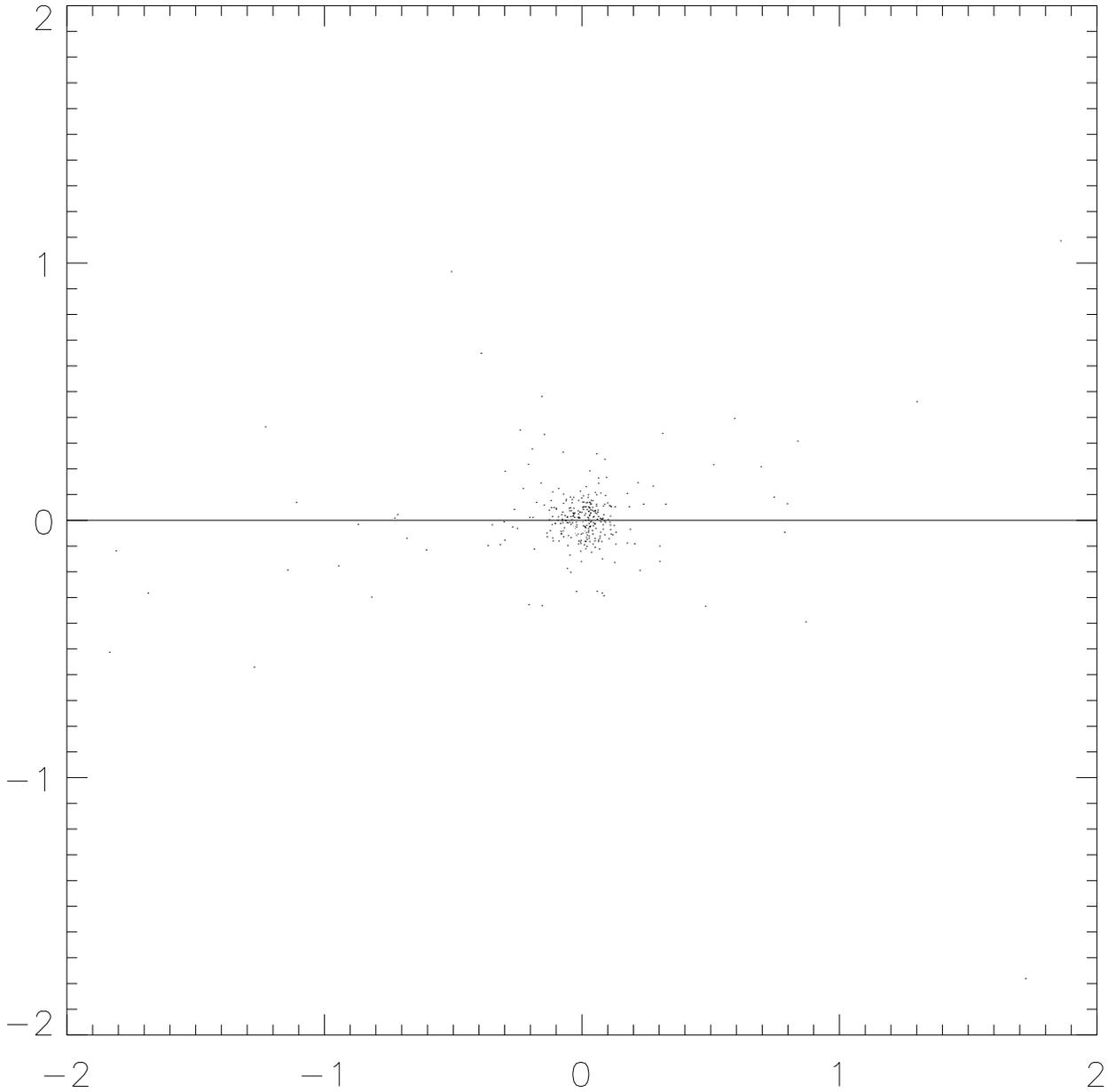}

\caption{\textbf{Snapshot near a maximum velocity dispersion}\label{fig:dsvelpic}
The dS stellar distribution as projected against the sky is shown here.
The plane of the orbit is indicated by the solid line. This view
corresponds to the dS when the velocity dispersion was approximately
10 times larger than its initial value and the core of the dS is apparent
even with a small number of stars in our N-body calculations.}
\end{figure}

%% file: PS/Film.tex
\begin{figure}
\epsscale{.86}
\plotone{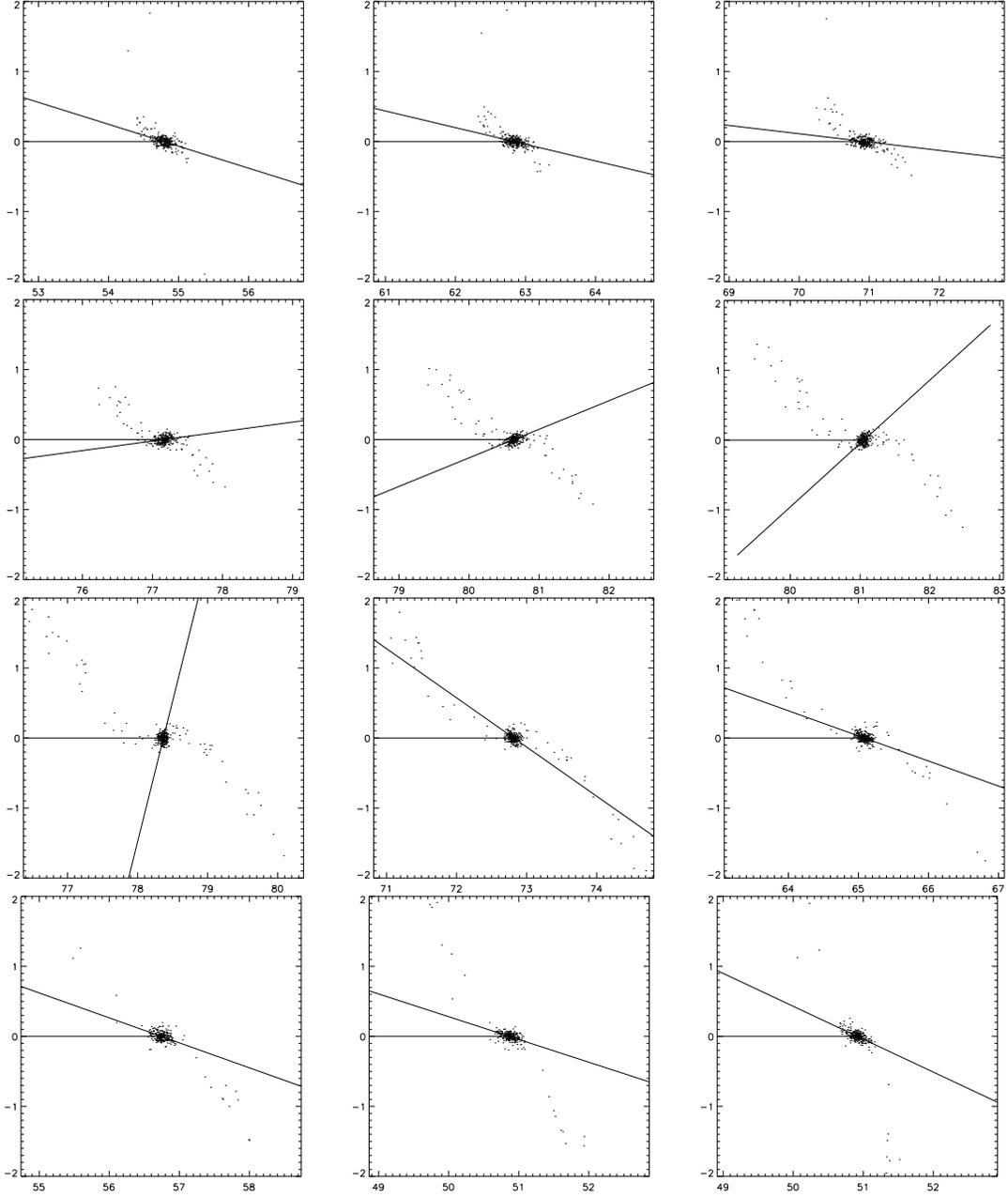}
\epsscale{1}
\caption[Snapshots of major axis rotation]{
\textbf{Snapshots of major axis rotation.}
\label{fig:turning:bar}\label{fig:snapshot}
Snapshots of the dS stars projected into the orbital plane are shown here.
Time advances from left to right and top to bottom with each panel 0.25 
time units apart. The horizontal line
represents the direction of the line-of-sight and the solid line
shows the direction of the major axis as computed from the moment
of inertia tensor of the stellar distribution. Rotation occurs from
snapshots 3 to 9. 
Note also that stellar ejection occurs primarily when the major axis begins
to turn, ejecting groups of stars into the tidal tails.}
\end{figure}

%% file: PS/order_Var_bon.tex
\begin{figure}
\plotone{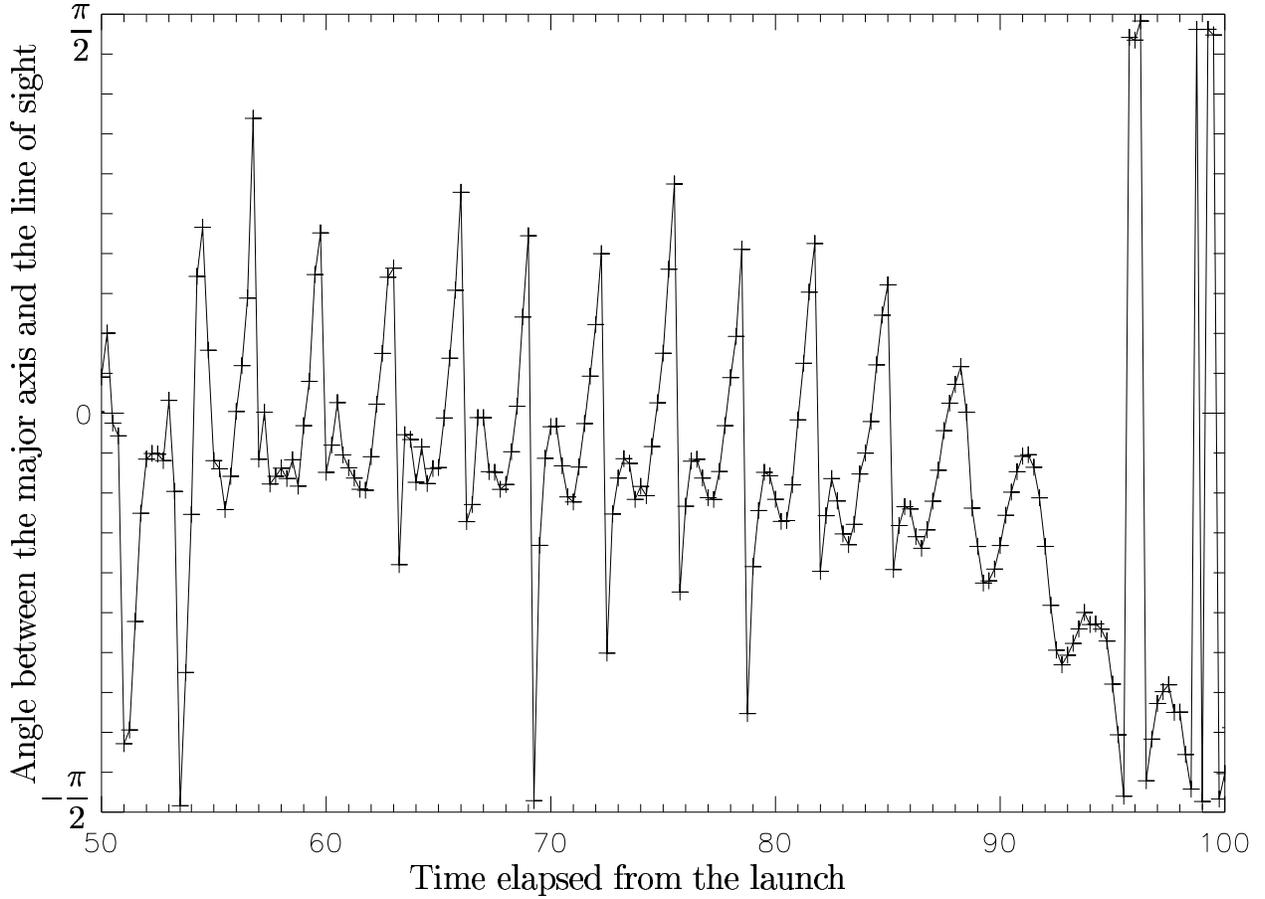}
\caption[Variation of the angle at first order in elliptical excitation]
{\textbf{Variation of the angle at first order in elliptical excitation.}
\label{fig:first:order}
Rotation of the dS bar appears as a discontinuity in the derived bar
angle. We have not 
represented the first $50$ time units where the major axis is not well 
defined and the angle varies chaotically.
Between successive perigalactica we see one discontinuity corresponding 
to half a turn of the \dS{}. This is characteristic of 
excitation at first order in $e$.}
\end{figure}

%% file: PS/dsprops.tex
\begin{figure}
\plotone{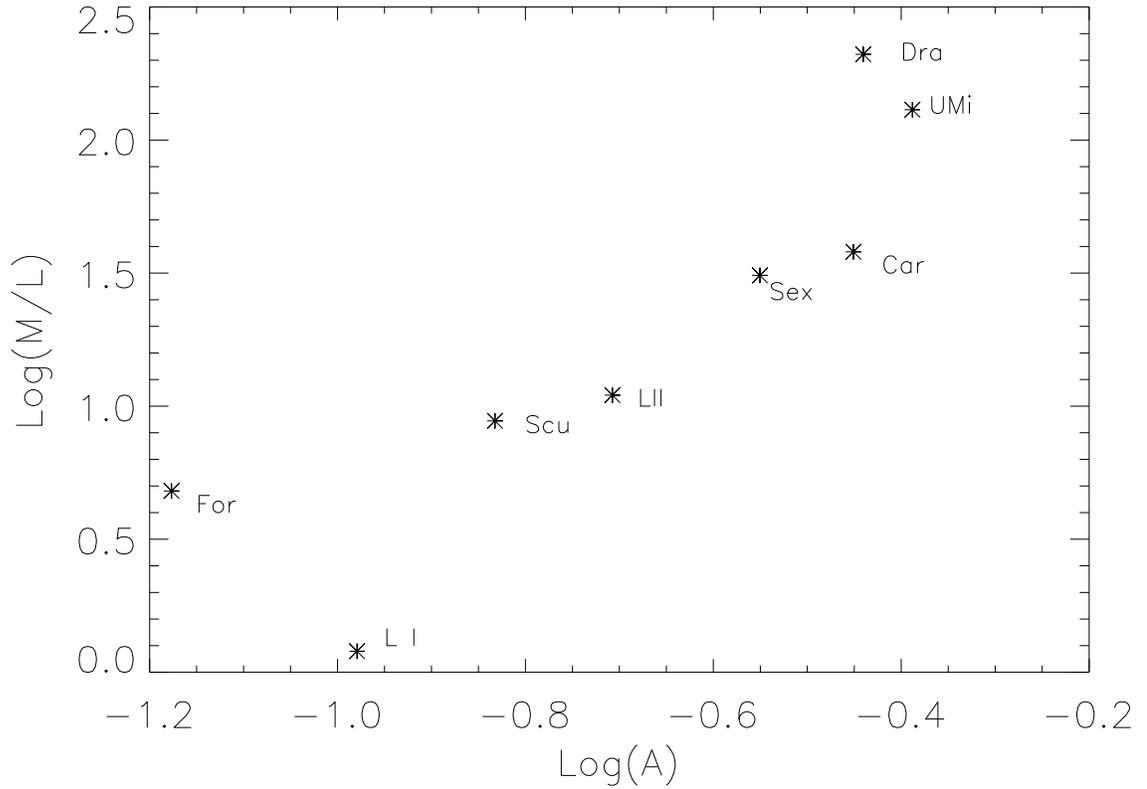}

\caption{\textbf{Mass/Light ratio against estimated growth rate.} 
\label{fig:dsprops} The Local Group dwarf spheroidal galaxy
virial ``Mass/Light'' ratio in solar units 
is plotted against a derived mode growth rate ($A=\omega_c /\omega_0 )$.
Parametric oscillations should inflate the velocity dispersion (or $M/L$)
with increasing $A$ as is seen here.}
\end{figure}